\pdfoutput=1
\documentclass[journal]{vgtc}                % final (journal style)
\ifpdf%                                % if we use pdflatex
  \pdfoutput=1\relax                   % create PDFs from pdfLaTeX
  \usepackage{graphicx}                % allow us to embed graphics files
  \DeclareGraphicsExtensions{.pdf,.png,.jpg,.jpeg} % for pdflatex we expect .pdf, .png, or .jpg files
\else%                                 % else we use pure latex
  \usepackage{graphicx}                % allow us to embed graphics files
  \DeclareGraphicsExtensions{.eps}     % for pure latex we expect eps files
\fi%

\graphicspath{{figures/}{pictures/}{images/}{./}} % where to search for the images

\usepackage{microtype}                 % use micro-typography (slightly more compact, better to read)
\PassOptionsToPackage{warn}{textcomp}  % to address font issues with \textrightarrow
\usepackage{textcomp}                  % use better special symbols
\usepackage{mathptmx}                  % use matching math font
\usepackage{times}                     % we use Times as the main font
\usepackage{cite}                      % needed to automatically sort the references
\usepackage{tabu}                      % only used for the table example
\usepackage{booktabs}                  % only used for the table example
\usepackage{amsmath}
\usepackage{amsfonts}
\usepackage{amsbsy}
\usepackage{times}
\usepackage{multirow}
\usepackage{makecell}
\usepackage{color}
\DeclareMathOperator*{\argmax}{arg\,max}
\DeclareMathOperator*{\argmin}{arg\,min}
\usepackage{xspace}
\makeatletter
\DeclareRobustCommand\onedot{\futurelet\@let@token\@onedot}
\def\@onedot{\ifx\@let@token.\else.\null\fi\xspace}
\def\eg{{e.g}\onedot} 
\def\ie{{i.e}\onedot} 

\onlineid{1075}

\vgtccategory{Research}
\vgtcpapertype{algorithm/technique}

\title{DRGraph: An Efficient Graph Layout Algorithm for Large-scale Graphs by Dimensionality Reduction}

\author{Minfeng Zhu, Wei Chen, Yuanzhe Hu, Yuxuan Hou, Liangjun Liu, and Kaiyuan Zhang}
\authorfooter{
\item
M. Zhu, W. Chen, Y. Hu, Y. Hou, L. Liu and K. Zhang are with State Key Lab of CAD\&CG, Zhejiang University.
E-mail: \{minfeng\_zhu, cadhyz, houyuxuan, liuliangjun\}@zju.edu.cn, chenwei@cad.zju.edu.cn, zhangkaiyuan20@gmail.com.
\item Wei Chen is the corresponding author.
}

\shortauthortitle{Biv \MakeLowercase{\textit{et al.}}: Global Illumination for Fun and Profit}
\abstract{
Efficient layout of large-scale graphs remains a challenging problem: the force-directed and dimensionality reduction-based methods suffer from high overhead for graph distance and gradient computation. In this paper, we present a new graph layout algorithm, called DRGraph, that enhances the nonlinear dimensionality reduction process with three schemes: approximating graph distances by means of a sparse distance matrix, estimating the gradient by using the negative sampling technique, and accelerating the optimization process through a multi-level layout scheme. DRGraph achieves a linear complexity for the computation and memory consumption, and scales up to large-scale graphs with millions of nodes. Experimental results and comparisons with state-of-the-art graph layout methods demonstrate that DRGraph can generate visually comparable layouts with a faster running time and a lower memory requirement.

} % end of abstract

\CCScatlist{ % not used in journal version
 \CCScat{K.6.1}{Management of Computing and Information Systems}%
{Project and People Management}{Life Cycle};
 \CCScat{K.7.m}{The Computing Profession}{Miscellaneous}{Ethics}
}

\vgtcinsertpkg

\begin{document}
\maketitle

\section{Introduction}\label{sec:introduction}
Graphs are common representations to encode relationships between entities in a wide range of domains, such as social networks~\cite{wu2016egoslider}, knowledge graph~\cite{wang2019kgat}, and deep learning~\cite{8019861}.
Node-link diagrams is an efficient method to depict the overall structure and reveal inter-node relations~\cite{kwon2018would}. 
Nevertheless, the layout influences the understanding of the graph.
For instance, it is typical to assume that two close nodes have high proximity even though no links exist among them \cite{mcgrath1996seeing}.
Therefore, preserving the neighborhood is an essential concept of graph layout.

Over the past 50 years, numerous efforts have been exerted on graph layout. 
However, an efficient graph layout algorithm remains a challenging problem for large-scale data.
Representatives include force-directed algorithms~\cite{fruchterman1991graph, kamada1989algorithm} and dimensionality reduction methods~\cite{kruiger2017graph}.
The force-directed methods solve the graph layout problem using a physical system with attractive and repulsive forces between nodes.
% The system energy of the system is minimized by moving nodes along the direction of the sum of forces.
Although force-directed methods are simple and easy to implement~\cite{battista1998graph}, they have a high computational complexity in calculating pairwise forces (the least one is $O(|{V}|\log|{V}|+|{E}|)$~\cite{hachul2004drawing}, where $|{V}|$ denotes the number of nodes and $|{E}|$ indicates the number of edges).
However, preserving distances between pairs of widely separated nodes may result in a large contribution to the cost function.
%In addition, force-directed algorithms treat local and global distances equally.
Thus, they are not good at preserving local structures \cite{maaten2008visualizing} and may converge to local minima and unpleasing results \cite{tamassia2013handbook}.

As an alternative, studies applied dimensionality reduction methods, such as multidimensional scaling (MDS) \cite{kruskal1964multidimensional}, principal component analysis (PCA) \cite{Jolliffe1986} and $t$-distributed stochastic neighbor embedding ($t$-SNE) \cite{maaten2008visualizing} for graph layout~\cite{harel2002graph,brandes2006eigensolver,kruiger2017graph}.
They usually minimize the difference between {the node similarity (\eg, the shortest path distance)} in the graph space and {the layout proximity (\eg, Euclidean distance)} in the layout space~\cite{yang2014optimization}.
Nonlinear dimensionality reduction methods aim to preserve the local neighborhood structure which is analogous to the concept of graph layout.
Though these methods can produce aesthetically pleasing results, they suffer from the high computational and memory complexity.
For instance, tsNET~\cite{kruiger2017graph} adopts $t$-SNE to capture local structures.
However, tsNET is amenable for graphs with only a few thousand nodes due to the following reasons: 
(1) the computational complexity of the shortest path distance is $O(|V|(|{V}|+|{E}|))$;
(2) pairwise {node similarities} require $O(|{V}|^2)$ computations;
(3) the gradient requires $O(|{V}|^2)$ distances for pairwise layout proximities.

In this paper, we propose a new graph visualization algorithm, DRGraph, that enhances the dimensionality reduction scheme to achieve the efficient layout of large graphs. 
Our approach differs from conventional dimensionality reduction algorithms in three aspects. 
First, we utilize a sparse graph distance matrix to simplify the computation of node similarities by only taking the shortest path distances between a node and its neighbors into consideration. 
Second, we employ the negative sampling technique~\cite{mikolov2013distributed} to compute the gradient on the basis of a subset of nodes, efficiently reducing the computational complexity.
Further, we present a multi-level process to accelerate the optimization process. 
By integrating these three techniques, DRGraph achieves a linear complexity for the computational and memory consumption (namely, $O(|{E}|+|{V}|+\text{TM}|V|)$ and $O(|{E}|+|{V}|)$, where ${\text{T}}$ denotes the number of iterations and $\text{M}$ is the number of negative samples).

We present a multi-threaded implementation of DRGraph and evaluate our DRGraph on various datasets quantitatively and qualitatively.
Generally, the single-thread version of DRGraph is roughly two times faster than GPU-accelerated tsNET while producing comparable layouts on moderate-sized graphs.
For large-sized graphs, DRGraph yields more expressive results, whereas tsNET cannot handle them without special optimizations. 
DRGraph runs at a comparable speed like $\text{FM}^3$~\cite{hachul2004drawing} and can preserve more topologically neighbors.
For the Flan\_1565 graph with 1,564,794 nodes and 56,300,289 edges, DRGraph consumes only 7 GB memory, whereas $\text{FM}^3$ requires approximately 44 GB memory.
Thus, DRGraph can easily scale up to large graphs with millions of nodes on commodity computers.
The source code of DRGraph is available at https://github.com/ZJUVAG/DRGraph.

%This paper is organized as follows. 
%Section \ref{sec:relatedwork} summarizes related works. 
%Section \ref{sec:method} presents our graph layout method.
%Experiments are introduced in Section \ref{sec:exp}, followed by discussions in Section \ref{sec:discussions} and conclusions in Section \ref{sec:conclusions}.

%, and scales up to large-scale graphs with millions of nodes on commodity computers.
%DRGraph compares favorably with $\text{FM}^3$~\cite{hachul2004drawing} in terms of quality and performance.
%In general, DRGraph is up to a thousand times faster than tsNET.
%DRGraph can generate comparable layouts to the tsNET on small graphs.
%The tsNET cannot handle large graphs without optimization acceleration, while DRGraph yields more intuitive visualizations on large-scale graphs.
%In addition, DRGraph also compares to $\text{FM}^3$~\cite{hachul2004drawing} in terms of quality and performance.
%DRGraph only consumes 7 GB memory to visualize a graph with 1,564,794 nodes and 56,300,289 links, while that of $\text{FM}^3$ is approximately 44 GB.
%and $\text{FM}^3$ needs approximately 58 GB.
%Compared to conventional solutions, DRGraph achieves up to ??? times acceleration.
%For instance, DRGraph is 1000 times faster than tsNET for a graph with 1,005 nodes and 3,808 links.
%DRGraph is 1.8 times as fast as , and 24.7 times as fast as Pivot MDS~\cite{brandes2006eigensolver} for a graph with 1,564,794 nodes and 56,300,289 links.
%The source code of DRGraph is available at https://github.com/DRGraphVIS/DRGraph.
\section{Related work}\label{sec:relatedwork}
\subsection{Dimensionality Reduction}%基于降维的方法
Dimensionality reduction methods convert a high-dimensional dataset into a low-dimensional space.
As a fundamental means for visualization, dimensionality reduction has been applied in a broad range of fields and ever-increasing datasets~\cite{7473883}.
%In recent decades, studies proposed numerous dimensionality reduction methods.
Classical techniques include PCA~\cite{Jolliffe1986}, Sammon mapping~\cite{1671271} and MDS~\cite{kruskal1964multidimensional}.
Researchers employ linear discriminant analysis \cite{fisher1936use} to reveal label information when data have associated class labels.
%LDA maximizes the distances between different clusters and minimizes the distances between data points in each cluster.
However, linear dimensionality reduction fails to detect nonlinear manifolds in high-dimensional space. 
Nonlinear dimensionality reduction algorithms aim to preserve local structures of nonlinear manifolds.
Isomap \cite{tenenbaum2000global} estimates the geodesic distance instead of Euclidean distance to minimize the pairwise distance error.

Recently, stochastic neighbor embedding (SNE) \cite{hinton2003stochastic} based approaches transform Euclidean distance into probability to measure similarities among the data points. 
$t$-distributed stochastic neighbor embedding ($t$-SNE) is proposed to solve the crowding problem \cite{maaten2008visualizing}. 
Though $t$-SNE shows its significant advantage in generating low-dimensional embedding, high computational complexity prevents it from being applied to large datasets.
%To solve this problem, researchers improve the efficiency of $t$-SNE by using tree-based method~\cite{kim2016pixelsne,van2014accelerating} and the negative sampling technique~\cite{mikolov2013distributed}. 
Barnes-Hut-SNE (BH-SNE) \cite{van2014accelerating} reduces the computational complexity from $O({N}^2)$ to $O({N}\log {N})$ by leveraging a tree-based method.
Tang et al.~\cite{tang2016visualizing} presented LargeVis to construct the $k$-nearest neighbor graph and accelerated optimization using the negative sampling technique~\cite{mikolov2013distributed}.
A-tSNE~\cite{7473883} progressively computes the approximated $k$-nearest neighborhoods and updates the embedding without restarting the optimization.
Nowadays, GPUs are widely employed for further acceleration \cite{chan2019gpu,pezzotti2019gpgpu}.
Though GPGPU-SNE \cite{pezzotti2019gpgpu} has a linear computational complexity, t-SNE-CUDA \cite{chan2019gpu} outperforms GPGPU-SNE due to the highly-optimized CUDA implementation.
Given the non-convexity of the objective function, $t$-SNE-based algorithms may end up in local minima and unpleasing layouts.
The multi-level concept has been widely used to address this problem \cite{arleo2018distributed}.
The multi-level representation is created by clustering \cite{7889042}, decomposition \cite{nguyen2017k}, anchor point \cite{joia2011local}, and Monte Carlo process (\eg, HSNE) \cite{pezzotti2016hierarchical}.
However, these methods suffer from high computation cost for generating the multi-level representation.
% In addition, the node or link weights of coarse graphs need to be carefully defined.
DRGraph introduces an enhanced multi-level scheme with a linear computational complexity.

\subsection{Graph Layout}
Graph layout algorithms map nodes of a given graph to 2D or 3D positions~\cite{liu2014survey}. 
The goal is to compute positions for all nodes according to the topological structure of the given graph.
Graph layout algorithms can be categorized into two classes, namely, force-directed and dimensionality reduction-based methods~\cite{gibson2013survey}.

\textbf{Force-directed methods.}
Most graph layout methods adopt the force-directed drawing algorithm because they are simple to understand and easy to implement.
There are two main classes of force-directed methods: spring-electrical and energy models~\cite{gibson2013survey}.

The spring-electrical model assigns attractive and repulsive forces between nodes.
The model moves each node along the direction of the composition force until the composition force on each node is zero.
Eades replaced nodes by steel rings and replaced edges with springs~\cite{eades1984heuristic}. 
To draw nodes evenly, Fruchterman and Reingold~\cite{fruchterman1991graph} modeled nodes as atomic particles and added repulsive forces between all nodes.
However, previous algorithms are time-consuming to visualize large-scale graphs due to high computational complexity.
Thus, previous studies employed simulated annealing techniques to optimize the spring-electrical model~\cite{frick1994fast,hu2005efficient}. 
%Cooling schedule gives better graph results. 
%Fruchterman et al.~\cite{fruchterman1991graph} proposed global temperature over all nodes to generate a more stable graph layout.
%Lower global temperature limits the maximum movement distance and consequences more stable graph layout. 
%Davidson et al.~\cite{davidson1996drawing} suggested an adaptive cooling schedule for different nodes. 
%Local temperature is calculated by detecting rotation and oscillation between last and current movement~\cite{frick1994fast}.
ForceAtlas2~\cite{jacomy2014forceatlas2} combines an adaptive-cooling schedule and a local temperature technique to produce continuous layouts.
To further expedite spring-electrical methods, the computational complexity of attractive and repulsive forces must be reduced. 
%For instance, the algorithm of Fruchterman and Reingold~\cite{fruchterman1991graph} computes $N^2$ repulsive forces between $N$ nodes in each iteration. 
Researchers adpot Barnes-Hut technique to accelerate the force calculation~\cite{hu2005efficient}. 
Multi-level method~\cite{hachul2004drawing,gajer2000grip,5742389} has been used widely in many graph layout methods. 
An initial layout is generated for the next larger graph that is drawn afterward~\cite{zinsmaier2012interactive,7889042}.
%These approaches aggregate subgraph as a unit to reduce the complexity greatly~\cite{walshaw2000multilevel,zinsmaier2012interactive,gajer2000grip,7889042}.
%Beck et al. \cite{beck2017taxonomy} provided a broad overview of dynamic graph visualization.
%Similarly, dynamic graph layout methods~\cite{6658746} compute the current graph layout with the initial layout derived from the previous frame~\cite{beck2017taxonomy}.

The energy model formulates the graph layout problem as an energy system and optimizes the system by searching a state with minimum energy~\cite{gansner2013maxent}.
A previous study generated a graph layout by solving the partial differential equations on the basis of the energy function~\cite{ren2017joint}. 
The concept of the KK algorithm proposed by Kamada and Kawai~\cite{kamada1989algorithm} is that Euclidean distances in the layout space should approximate graph-theoretic distances, \ie, the shortest path distance. 
Incremental methods~\cite{cohen1997drawing} accelerate the optimization by arranging a small portion of the graph before arranging the rest.
Stress majorization is employed to improve the computation speed and graph layout quality~\cite{gansner2004graph}.
Stress function can be reformulated to draw graphs with various constraints~\cite{wang2018revisiting,tim2006stress,hoffswell2018setcola}, including length, non-overlap, and orthogonal constraints~\cite{7192690,7192733,ulf2014stress}.
Pivot MDS~\cite{brandes2006eigensolver} first places anchor nodes and then locates other nodes on the basis of their distances to anchor nodes.

\textbf{Dimensionality reduction based methods.}
Graph layout by dimensionality reduction aims to preserve graph structures. 
Utilizing dimensionality reduction techniques to study graph layout requires further exploration. 
Recent works~\cite{yang2014optimization,lu2016doubly,kruiger2017graph} pursued this line of thought and illustrated how to use dimensionality reduction for graph layout. Graph layout by dimensionality reduction can be classified into projection and distance-based methods.

Projection-based methods have two stages: first, embed graph nodes into a high-dimensional space and then project vectors into low-dimensional space.
High-dimensional embedding (HDE)~\cite{harel2002graph} adopts PCA to project the graph.
Koren et al.~\cite{koren2004graph} improved HDE by replacing PCA with subspace optimization. 
Zaor{\'a}lek et al.~\cite{zaoralek2014dimension} compared several different dimensionality reduction methods for graph layout.
More recently, powerful deep neural networks are also utilized to learn how to draw a graph from training examples~\cite{wang2019deepdrawing,kwon2019deep}.
However, the pairwise similarity loss of deep-learning methods commonly has a quadratic computational complexity with respect to the number of nodes.
Distance-based methods adopt graph-theoretic distance instead of the distance in high-dimensional space used by projection-based methods.
s-SNE~\cite{lu2016doubly} is developed by considering spherical embedding and resolves the "crowding problem" by eliminating the discrepancy between the center and the periphery.
tsNET~\cite{kruiger2017graph} utilizes neighborhood-preserving $t$-SNE technique for graph layout.
Dimensionality reduction approaches with high efficiency can be employed to accelerate tsNET.
The single-thread version of DRGraph is faster than tsNET accelerated by the t-SNE-CUDA algorithm \cite{chan2019gpu}.
DRGraph optimizes the objective function with the negative sampling technique~\cite{mikolov2013distributed} which reduces the computational complexity to linear. Also, we employed an efficient multi-level representation to propagate gradient information and draw graphs from coarse to fine.

\section{Method}\label{sec:method}
% We introduce the framework of tsNET~\cite{kruiger2017graph} in Section \ref{section:tsnet} and explain our new solution in Section \ref{section:drgraph}. The advantages of DRGraph over tsNET are discussed in Section \ref{section:comparisons}.

\subsection{Background on Graph Layout with tsNET}\label{section:tsnet}
Our approach takes a similar framework as that of tsNET \cite{kruiger2017graph}. 
% Here we describe tsNET with a set of notations listed in Table \ref{tab:notations}.
Formally, let $G = (V, E)$ be an undirected unweighted graph with a set of nodes $V=\{v_1, v_2,..., v_{|V|}\}$ and a set of edges $E=\{e_1, e_2,..., e_{|E|}\}$. 
Each edge is a connection between two nodes: $e=(v_i,v_j) \in V \times V $.
Then, the graph layout problem is formulated as embedding a given graph to 2D or 3D space: $\phi: G \to Y, Y=\{y_1,y_2,...,y_{|V|}\}$, where $y_i$ is the layout position of node $v_i$.

% \begin{table}[!b]
%   \centering
%     \caption{Mathematical Notations}
%   \begin{tabular}{p{0.2\columnwidth}p{0.7\columnwidth}} % Column formatting, @{} suppresses leading/trailing space
%      \toprule
%      Symbol & Description\\
%      \midrule
%      $G=(V,E,A)$ & A graph\\
%      $V$ & The node set of graph $G$\\
%      $E$ & The link set of graph $G$\\
%      $A$ & The adjacent matrix of graph $G$\\
%      $Y$ & The layout positions of nodes of graph $G$\\
%      $GD$ & The graph distance matrix\\
%      $LD$ & The layout distance matrix\\
%      $NS$ & The node similarity matrix\\
%      $LP$ & The layout proximity matrix\\
%      $D(\cdot,\cdot)$ & The loss function\\
%      $D_{KL}(\cdot||\cdot)$ & The Kullback-Leibler divergence\\
%      $SPD(v_i,v_j)$ & The shortest path distance between $v_i$ and $v_j$ \\
%      $NNG(v_i,k)$ & The $k$-order nearest neighbors of $v_i$ in graph space\\
%      $NNL(y_i,k)$ & The $k$-nearest neighbors of $y_i$ in layout space\\
%      \bottomrule
%   \end{tabular}
%   \label{tab:notations}
% \end{table}

Graph layout methods are tied by an optimization problem~\cite{yang2014optimization} that minimizes the difference between the graph space and the layout space.
The \textit{node similarity} ($NS$) is defined as the pairwise similarity between two nodes in the graph space.
The \textit{layout proximity} ($LP$) is defined as the pairwise proximity between two nodes in the layout space.
Connected nodes with high node similarity should preserve high layout proximity in the layout space.
A \textit{loss function} $D(NS, LP)$ models the difference between the node similarity and the layout proximity.
The Kullback-Leibler divergence $D_{KL}(\cdot||\cdot)$ formulates the graph layout problem as optimizing the following objective function:
\begin{eqnarray}
Y^* &=& \argmin_Y D_{KL}(NS||LP) \nonumber\\
%&=& \argmin_Y \sum_{i\not=j}NS_{ij}\log \frac{NS_{ij}}{LP_{ij}}\nonumber \\
&=& \argmin_Y \sum NS_{ij}\log NS_{ij} - NS_{ij}\log LP_{ij},
\end{eqnarray}
where $NS_{ij}$ is the node similarity between $v_i$ and $v_j$, and $LP_{ij}$ is the layout proximity between $y_i$ and $y_j$, and $Y^*$ is the optimal graph layout. 
Given that the first term is a constant, the problem is equivalent to the following optimization problem:
\begin{eqnarray}
Y^* = \argmax_Y \sum \quad NS_{ij}\log LP_{ij}.%_{i\not=j}
\end{eqnarray}

\textbf{The node similarity} is computed by graph-theoretic distance $GD$ in the graph space.
We compute a graph distance matrix by leveraging the shortest path distance (SPD) using a breadth-first search: $GD_{ij} = SPD(v_i,v_j)$.
$SPD(v_i,v_j)$ is the shortest path distance between nodes $v_i$ and $v_j$.
The node similarity matrix $NS$ is given by transforming the graph distance using a {similarity function} (\eg, Gaussian distribution):
\begin{eqnarray}
NS_{j|i}&=&\frac{exp(-GD_{ij}^2/2\sigma_i^2)}{\sum_{k\not=i}exp(-GD_{ik}^2/2\sigma_i^2)},~NS_{i|i}=0,\label{eq:NS}\\
NS_{ij}&=&(NS_{i|j}+NS_{j|i})/(2|V|),
\end{eqnarray}
where $\sigma_i$ is the variance of Gaussian distribution on node $v_i$.

\textbf{The layout proximity} is measured by the layout distance $LD$ between nodes' positions in the layout space.
We can compute {layout distance $LD$ using Euclidean distance}:
$LD_{ij}=\|y_i-y_j\|_2$.
Then, the {layout proximity $LP$ is measured by a proximity function} (\eg, Student's $t$-distribution).
The proximity function captures important locality properties in the layout space, providing an appropriate scale to connect the node similarity and the layout proximity.
The layout proximity of the pair $(y_i,y_j)$ in the layout space can be formulated as follows: 
\begin{eqnarray}
LP_{ij}=\frac{(1+LD_{ij}^{2b})^{-1}}{\sum_{k\not=l}(1+LD_{kl}^{2b})^{-1}},
\end{eqnarray}
where $LP_{ij}$ denotes the layout proximity, $LD_{ij}$ denotes the layout distance between $v_i$ and $v_j$, and $b$ is a parameter to control the shape of the distribution.
When $b=1$, $LP$ is equivalent to the normalized Student's $t$-distribution (a single degree of freedom) used by tsNET.

tsNET modifies the objective function with two additional cost terms, and tsNET* further assigns initial values by Pivot MDS (PMDS) \cite{brandes2006eigensolver}.
The tsNET algorithm is useful for neighborhood preservation.
However, tsNET is amenable for graph data with only a few thousand nodes due to the following reasons.
First, tsNET must measure graph distances between all node pairs to construct node similarity.
All pairwise {shortest path distances} require $O(|{V}|(|{V}|+|{E}|))$ computations using the breadth-first search. 
Second, computing the node similarity needs $O(|{V}|^2)$ computations, because computing the normalization terms needs to sum over $|{V}|^2$ graph distances.
Third, the gradients require $|{V}|^2$ pairs of Euclidean distances in each iteration.
Thus, tsNET has a quadratic computational and memory complexity:
\begin{align}
&{C}^{{computation}}_{{tsNET}}=O(|{V}|(|{V}|+|{E}|)+|{V}|^2+\text{T}|{V}|^2)\label{eq:1}\\
&{C}^{{memory}}_{{tsNET}}=O(|{V}|^2)\label{eq:2}
\end{align} 
where T is the number of iterations.

\subsection{DRGraph}
\label{section:drgraph}
\begin{figure}[!t]
  \centering
  \includegraphics[width=0.99\columnwidth]{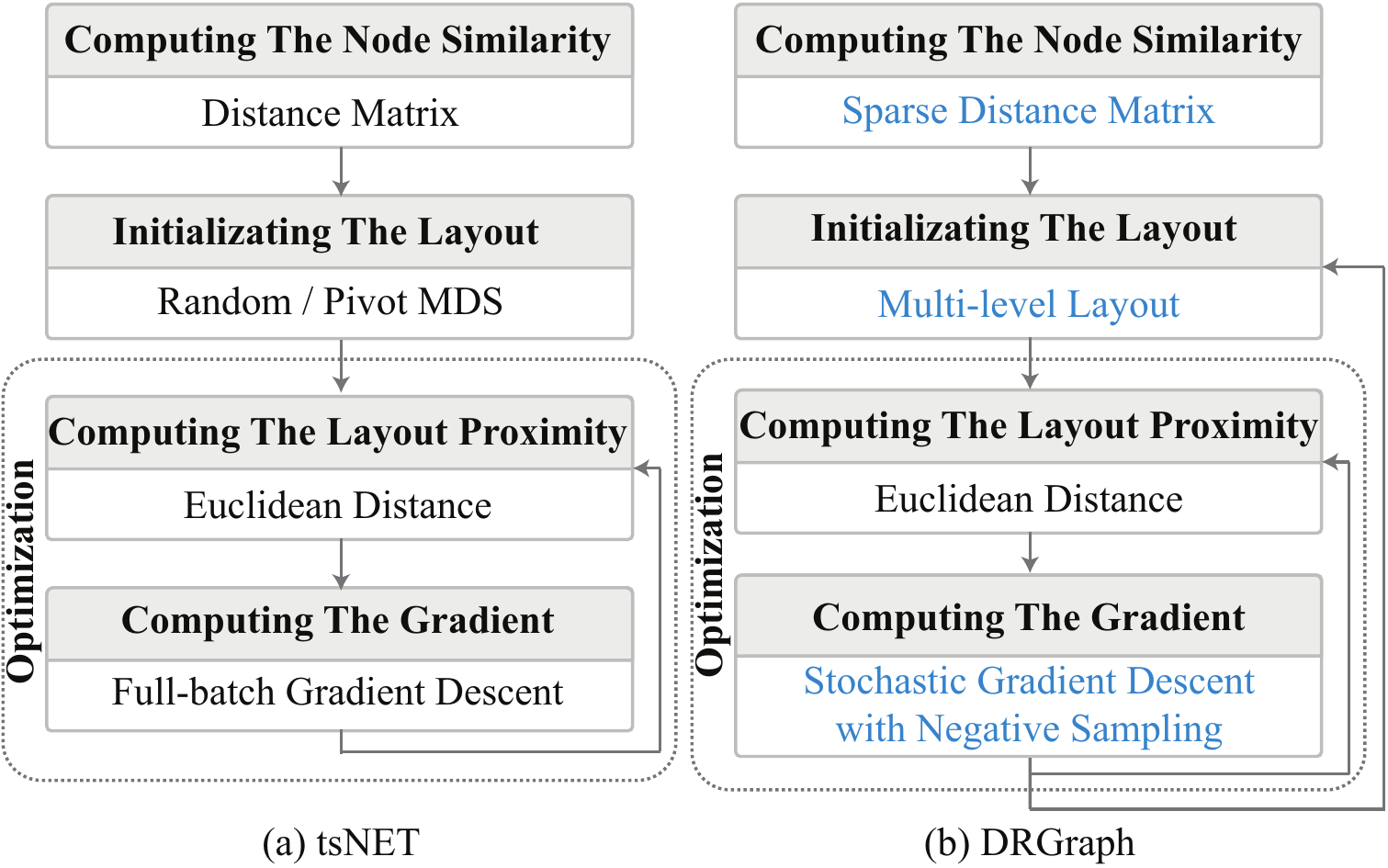}
  \caption{Both tsNET and DRGraph employ an iterative process to optimize the layout with respect to the layout proximity. 
  DRGraph utilizes a sparse distance matrix, a multi-level scheme, and a negative sampling technique to accelerate the layout process.
  }
  \label{figure:framework}
\end{figure}

We seek to overcome the performance overhead of tsNET in three aspects. 
Particularly, our approach utilizes a sparse graph distance matrix to simplify the computation of pairwise node similarities, the negative sampling technique to approximate the gradient on the basis of a subset of nodes and a multi-level process to accelerate the optimization process. 
By integrating them into a new pipeline, called DRGraph, a linear complexity for the computation and memory consumption is achieved.
Figure \ref{figure:framework} compares the framework of tsNET and DRGraph.
In addition to the new layout pipeline, three new components are highlighted in blue font, namely, sparse distance matrix, multi-level layout, and negative sampling. The details are elaborated as follows.

\subsubsection{Sparse Distance Matrix}
\begin{figure}[!t]
  \centering
  \includegraphics[width=0.99\columnwidth]{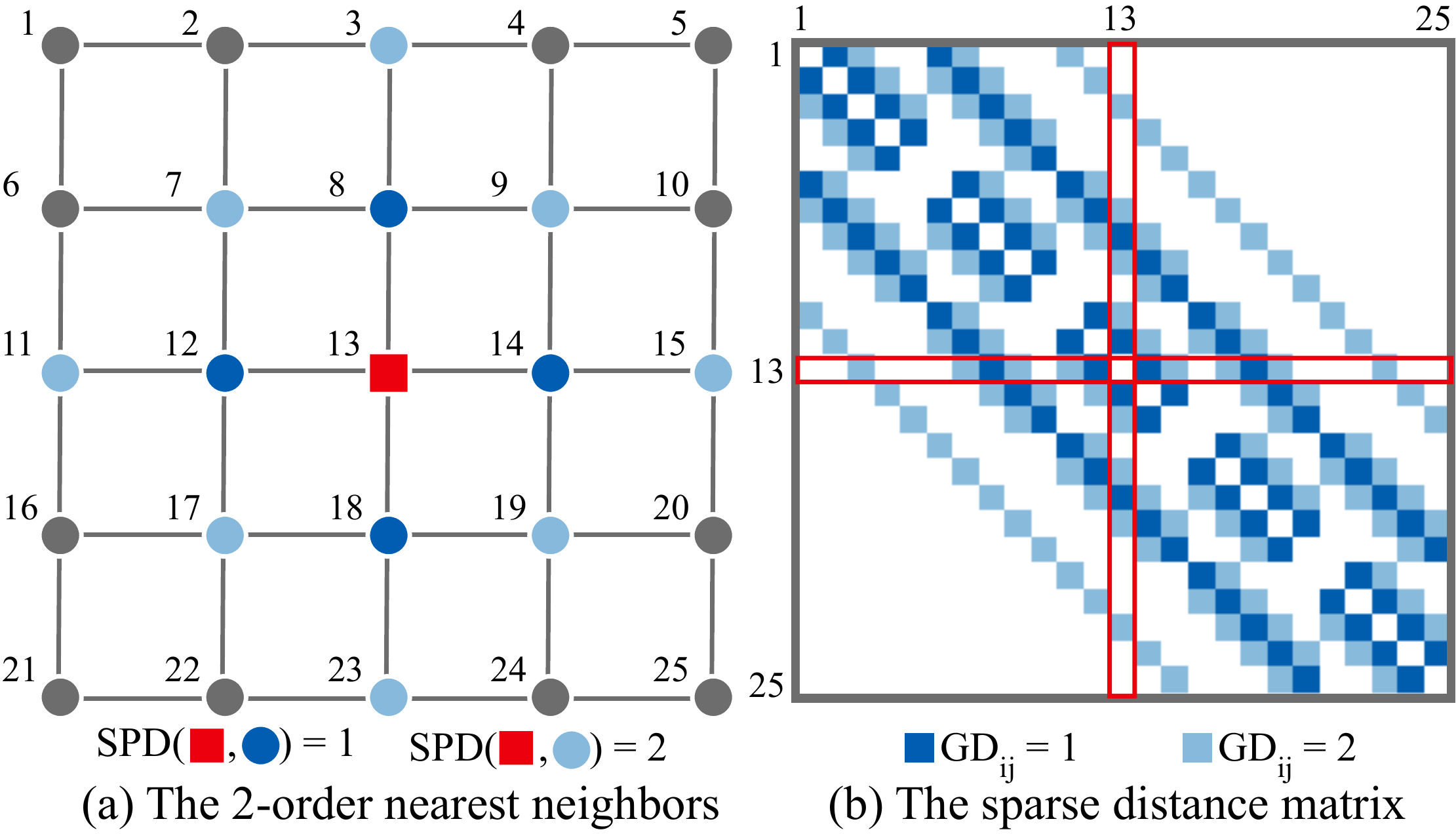}
  \caption{(a) The $2$-order nearest neighbors of the 13th node in a graph.
  (b) The sparse distance matrix constructed on the $2$-order nearest neighbors.
  $SPD$ denotes the shortest path distance in the graph space and $GD$ denotes the graph distance matrix.
  }
  \label{figure:sparsematrix}
\end{figure}

To reduce the computation and memory requirements of {the node similarity}, we propose to approximate the node similarity using a sparse representation without a significant effect on the layout quality.
This scheme works due to the following observations. 
First, the node similarity of two nearby nodes with a small shortest path distance is relatively high according to the definition (Eq. \ref{eq:NS}).
Second, the node similarity between widely separated nodes is almost infinitesimal.
Therefore, a small shortest path distance has a significant contribution to the objective function.
We utilize a sparse distance matrix to simplify the computation of {pairwise node similarities} by using the shortest path distance between a node and its neighbors (see Figure \ref{figure:sparsematrix}).

We define $k$-order nearest neighbors $NNG(v_i,k)$ of node $v_i$ as a set of nodes whose shortest path distances to $v_i$ are less than or equal to $k$:
\begin{equation}
    NNG(v_i,k) = \{v| SPD(v_i,v) \leq k, v \neq v_i, v \in V \}.
\end{equation}
For instance, first-order nearest neighbors $NNG(v_i,1)$ is the set of nodes connected to $v_i$.
We can compute a sparse distance matrix where $GD_{ij}=SPD(v_i,v_j)$ if $v_j \in NNG(v_i,k)$.
%The node similarity $NS_{ij}$ is reformulated as: 
Eq. \ref{eq:NS} is reformulated as:
\begin{equation}
    N\!S_{j|i}\!=\!\left\{
\begin{aligned}
&\frac{exp(-GD_{ij}^2/2\sigma_i^2)}{\sum\limits_{v_l \in \!N\!N\!G(v_i,k) }exp(-GD_{il}^2/2\sigma_i^2)},\text{if}~v_j\!\in\!N\!N\!G(v_i,k)\\
&0,\text{otherwise}\\
\end{aligned}
\right.
\end{equation}

The worst case of finding a node's neighborhoods is exploring all edges in $O(|E|)$.
To find the $k$-order nearest neighbors, a breadth-first search must access $\min(|E|,(|E|/|V|)^k)$ nodes, where $|{E}|/|{V}|$ denotes the average degree \cite{russell2016artificial}.
Therefore, we can generate a sparse graph distance matrix $GD$ by finding the $k$-order nearest neighbor set in $O(\text{D}|{V}|),~\text{D}=\min(|E|,(|E|/|V|)^k)$.
Measuring and storing node similarity with a large nearest neighbor set for graphs with millions of nodes is infeasible due to the memory limitation. 
In most instances, first-order nearest neighbors are sufficient to capture neighborhood information and produce pleasing graph layouts.
We can generate $GD$ in $O(|E|)$ time by leveraging the first-order nearest neighbor set.

For node similarity $NS$, the value of the graph distances $GD_{ij}$ ranges from 1 to $k$.
We can pre-compute the Gaussian distribution for $k$ different values and measure $NS$ with $O(k|V|)$ calculations.
Thus, computing the node similarity using the sparse graph distance matrix has a $O(k|V|)$ computational complexity. 
In particular, the node similarity can be measured in $O(|V|)$ if we employ the first-order nearest neighbors.

\subsubsection{Negative Sampling}
We employ the negative sampling technique~\cite{mikolov2013distributed} to approximate the gradient using a small set of nodes.
We sample one positive node and M negative nodes for each gradient computation.
We use a logistic regression to separate one positive node $y_j$ and M negative nodes $y_{j_m},m=1,...,\text{M}$. 
The likelihood function is defined as follows:
\begin{equation}
  \log LP_{ij} + {\sum}^{M}_{m=1}\gamma\log (1-LP_{ij_m}),
\end{equation}
in which $\gamma$ is a weight assigned to the negative samples.
In this way, the gradient of each node needs $\text{M}+1$ Euclidean distances.
We randomly sample the positive node $y_j$ on the basis of the edge probability $NS_{ij}$.
We identify the negative sample $y_{j_m}$ according to the node weight $\sum_l NS_{j_m,l}$.
We reformulate the optimization problem as follows:
\begin{eqnarray}\label{eq:objectfunction}
D(NS||LP)&\!=\!&
\mathop{\mathbb{E}}\limits_{(i,j) \sim NS_{ij}}[\log LP_{ij}+\sum_{m=1}^\text{M} \gamma \log(1-LP_{ij_m})].
\end{eqnarray}
The partial derivative of the objective function (Eq. \ref{eq:objectfunction}) is derived as:
\begin{equation}
\label{eq:gradient}
\frac{\partial D}{\partial y_i} = \underbrace{-\frac{2b(y_i-y_j)^{2b-1}}{1+LD_{ij}^{2b}}}_{\text{attractive force}} + \sum\limits_{m=1}^M \underbrace{\gamma \frac{2b(y_i-y_{j_m})}{LD_{ij_m}^2(1+LD_{ij_m}^{2b})}}_{\text{repulsive force}}.
\end{equation}
The gradient shows that each node receives one attractive force and M repulsive forces.
%To solve the optimization problem, we randomly select two vertices and samples M negative points ($y_{j_k}$) by adopting the negative sampling technique used in LargeVis.
During optimization, we randomly select a node and compute the gradient of the node. 
Each gradient computation takes $O(\text{M})$ time, where $\text{M}$ is the number of negative samples.
Consequently, the negative sampling technique reduces distance calculation from $O(|V|)$ to $O(\text{M})$ for the gradient computation of each node.
We define one iteration as computing gradient $|V|$ times.%, which means we update the position of all nodes once.
In practice, we find that the number of iterations is usually a constant.
The computational complexity of the optimization is $O(\text{T}\text{M}|V|)$, where $\text{T}$ denotes the iteration number.
Therefore, the objective function can be effectively optimized by the stochastic gradient descent algorithm in linear time.

\subsubsection{Multi-level Layout Scheme}
The multi-level approach has been used widely in many graph layout methods~\cite{gajer2000grip,hachul2004drawing}. 
It starts from a coarse graph layout and iteratively optimizes to a refined layout. 
We design a multi-level scheme to generate a multi-level representation in linear time.
Our scheme comprises three steps: coarsening, layout of the coarsest graph, and refinement.

\begin{figure}[!ht]
  \centering
  \includegraphics[width=0.99\columnwidth]{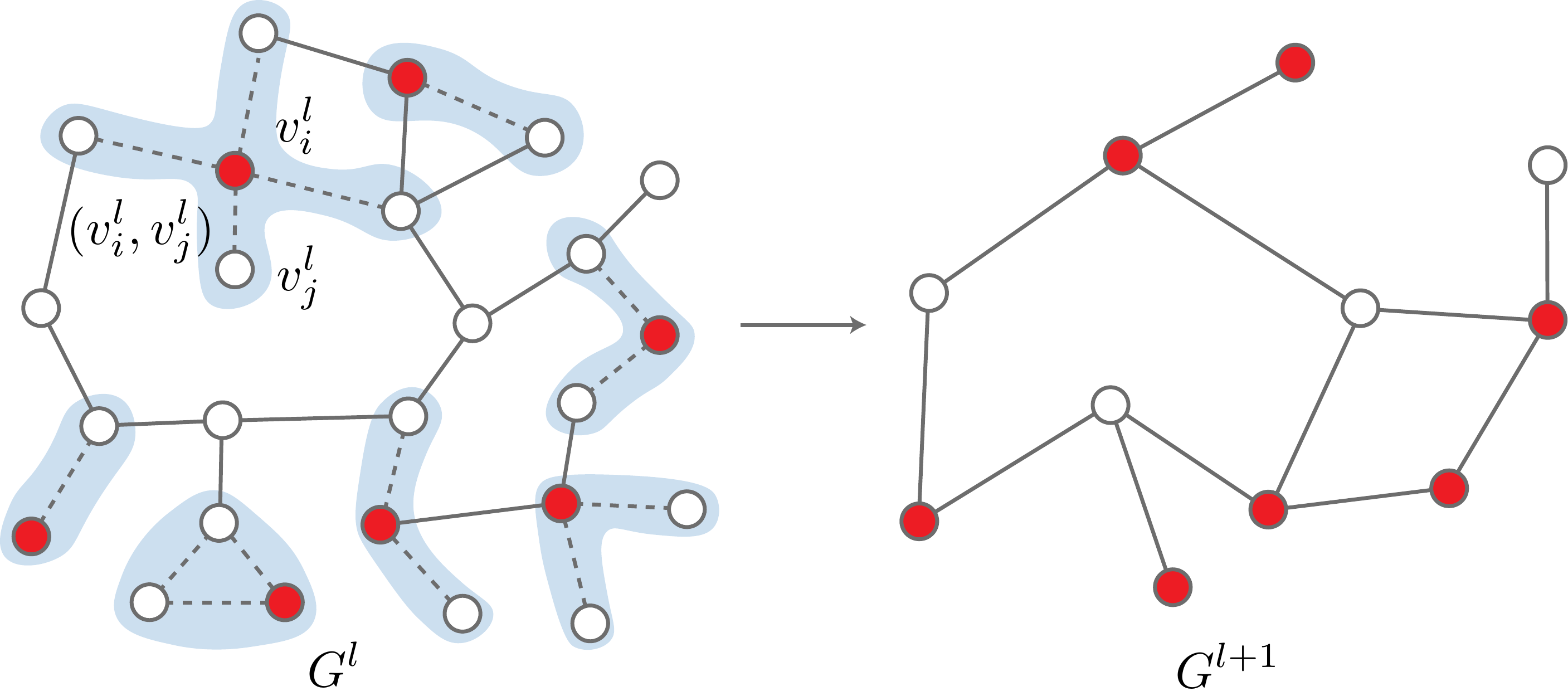}
  \caption{An example of graph coarsening. We randomly select nodes (red circles) of graph $G^l$ and group their first-order neighbors (covered by blue regions) into new nodes of $G^{l+1}$.}
  \label{figure:coarsening}
\end{figure}

\textbf{Coarsening.}
The coarsening step generates a series of coarse graphs $G^0, G^1, G^2,..., G^L$ with decreasing sizes, where $G^0$ is the original graph and $G^L$ is the coarsest graph.
Given a graph $G^l=(V^l, E^l)$, we generate a coarser graph $G^{l+1}$ as follows.
%we partition the node set of $G^l$ into disjoint subsets and generate a coarser graph $G^{l+1}$.
%Each subset represents a new node in the graph $G^{l+1}$.
First, we randomly select a node $v^l_i$ (red nodes in Figure \ref{figure:coarsening}). 
Then, we assign $v^l$ and its first-order neighbors (nodes in blue regions) into a new node in $G^{l+1}$.
%We denote that $v^l_i$ belongs to $v^{l+1}_j$ if $v^l_i$ is assigned to $v^{l+1}_j$.
Third, we delete edge $(v^l_i,v^l_j)\in E^l$ in the graph $G^l$, if $v^l_i$ and $v^l_j$ are assigned into the same node in $G^{l+1}$.
This process is repeated until no nodes can be assigned.
The coarsening step reduces the number of nodes at each level.
However, when the size of $G^{l+1}$ is very close to $G^{l}$, the cost of the multi-level algorithm significantly increases~\cite{hachul2004drawing} and the size of the coarsest graphs can not be further reduced.
Therefore, it is sufficient to cease the coarsening process when $|V^{l+1}|>\rho|V^{l}|$.
We choose $\rho$ to be 0.8 to achieve the balance between the computational efficiency and the global structure extraction \cite{hu2005efficient}.

\textbf{Layout of the coarsest graph.}
For the coarsest graph $G^L$, we layout the graph with random initialization.
The optimal layout of the coarsest graph can be found at a low cost.

\textbf{Refinement.}
Once we generate the graph layout of a coarse graph $G^{l+1}$, the initial layout of the finer graph $G^{l}$ is derived from $G^{l+1}$.
We set the position of a node $v^{l}$ in $G^{l}$ to be the position of node $v^{l+1}$ in $G^{l+1}$, if $v^{l}$ is assigned to $v^{l+1}$ in the graph coarsening step.
Then, we recursively refine the layout until we complete the finest graph $G^0$.

Conventional multi-level techniques require computing new node similarities for each level if we draw each graph level individually.
We optimize the graph layout jointly by sharing the gradient through the multi-level representation.
We pre-compute the node similarity between nodes of $G^0$ just once before the optimization.
For the layout of $G^l$, we select node $v_i^0$ from $G^0$ and compute the gradient of node $v_i^0$. 
We forward the computed gradient of node $v_i^0$ to node $v_j^l$ in $G^l$ if $v_i^0$ is assigned to $v_j^l$ in the graph coarsening step.

The running time of the multi-level scheme denotes the time of creating a series of coarse graphs: $\sum_{l=1}^Lt_{create}(G^l)$. The worst case of creating $G^l$ from $G^{l-1}$ is accessing all nodes and edges in $O(|V^l|+|E^l|)$. 
Let us assume that $|V^l|\leq 0.8|V^{l-1}|$ and $|E^l|\leq 0.8|E^{l-1}|$ for all $l=1,...,L$, then $\sum_{l=1}^Lt_{create}(G^l) = (|V| + |E|)(1+\frac{4}{5}+...+(\frac{4}{5})^L)\leq 5(|E| + |V|)$.
The computational complexity is linear in $O(|V|+|E|)$.
For memory complexity, we need to store the node index of all coarse graphs.
Each graph $G^l=(V^l,E^l)$ needs $O(|V^l|+|E^l|)$ space. 
With the assumption of $|V^l|\leq 0.8|V^{l-1}|$ and $|E^l|\leq 0.8|E^{l-1}|$, the multi-level approach yields a linear memory complexity of $O(|V|+|E|)$.

\subsubsection{Complexity Analysis}
\textbf{Computational complexity.}
The computational complexity of our algorithm includes $k$-order nearest neighbor set searching $O(\text{D}|V|)$, node similarity computation $O(k|V|)$, coarse graph generation $O(|V|+|E|)$ and optimization $O(\text{TM}|V|)$.
The total computational complexity of DRGraph is derived as:
\begin{equation}
{C}^{{computation}}_{{DRGraph}}=O(D|V|+k|V|+|E|+\text{TM}|V|),\label{eq:3}
\end{equation} 
where $\text{D}=\min(|{V}|,(|{E}|/|{V}|)^k)$.
DRGraph achieves a linear computational complexity of $O(|E|+|V|+\text{TM}|V|)$ if we employ the first-order nearest neighbor set ($k$=1).

\noindent\textbf{Memory consumption.}
DRGraph requires $(D|V|)$ memory to store the similarity of sparse nodes, $O(|V|+|E|)$ memory to record all coarse graphs, and $O(|V|)$ memory to store the layout position of nodes.
The total memory consumption of DRGraph is derived as:
\begin{equation}
{C}^{{memory}}_{{DRGraph}}=O(\text{D}|V|+|V|+|E|)\label{eq:4}.
\end{equation} 
We can reduce the memory complexity of DRGraph to $O(|E|+|V|)$ if we employ the first-order nearest neighbor set ($k$=1).

\subsection{DRGraph versus tsNET}
\label{section:comparisons}

One limitation of tsNET is that its computational and memory complexities are quadratically proportional to the graph size.
Contrastingly, DRGraph yields a linear computational complexity and only requires a linear memory consumption to store similarity and coarse graphs.
%The detailed complexity comparison can be shown in Eqs. \ref{eq:1}, \ref{eq:2}, \ref{eq:3} and \ref{eq:4}. 

DRGraph and tsNET are not guaranteed to converge to the global optimum due to the non-convexity objective function of $t$-SNE.
This not only needs to modulate several parameters but also easily converges to local minima.
In addition, a different random initialization may lead to a different graph layout.
Though tsNET* initializes the layout using PMDS, the result remains unpleasing if PMDS fails to maintain the global structure given a small number of pivots.
DRGraph adopts the multi-level scheme to coarsen graphs and capture the global structure progressively.
DRGraph can find the optimal initial layout using the coarsest graph.
Besides, tsNET may distort the PMDS layout since PMDS preserves short and long shortest-path distances, which conflicts with the neighbor-preserving nature of tsNET.
DRGraph successively refines graph layouts from the coarsest to the original one, resulting in no distortions between coarse graphs.

% Thus, DRGraph produces a visually aesthetic result.
%find a better initial layout and produce a visually aesthetic result.
%for coarse graphs efficiently.

% In summary, DRGraph outperforms tsNET in terms of the time and memory consumptions while producing a comparable quality.

%Generally, it is acceptable if we increase the number of iterations to compute a good layout for the coarsest graph.

%A drawback of DRGraph is that the estimated gradient do not have any error bound due to the random selection of the negative sampling technique.
%Because the objective function of DRGraph is non-convex, the final graph layout is strongly influenced by the coarsest graph layout.

\section{Results}\label{sec:exp}
In this section, we evaluate the efficiency and effectiveness of DRGraph.
We conduct all experiments on a desktop PC with Intel(R) Core(TM) i7-6700 CPU, 64 GB memory, and Ubuntu 16.04 installed.

\textbf{Datasets.}
We perform experiments on a broad range of datasets selected from the University of Florida Sparse Matrix Collection~\cite{Davis:2011} and tsNET\cite{kruiger2017graph} (Table \ref{table:dataset}).

\begin{table}[!ht]\setlength{\tabcolsep}{6.5pt}
\caption{Test datasets}
\label{table:dataset}
\scalebox{0.9}{
\begin{tabular}{@{}lrrl@{}}
\toprule
{Dataset}   & {\#Nodes} & {\#Edges} & {Description} \\ \midrule
dwt\_72 & 72 & 75 & planar structure\\
lesmis & 77 & 254 & collaboration network \\
can\_96 & 96 & 336 & mesh \\
rajat11 & 135 & 377 & miscellaneous network\\
jazz & 198 & 2,742 & collaboration network \\
visbrazil & 222 & 336 & tree-like network \\
grid17 & 289 & 544 & grid \\
mesh3e1 & 289 & 800 & grid \\
netscience & 379 & 914 & collaboration network \\
dwt\_419 & 419 & 1,572 & planar structure \\
price\_1000 & 1,000 & 999 & tree-like network \\
dwt\_1005 & 1,005 & 3,808 & planar structure \\
cage8 & 1,015 & 4,994 & miscellaneous network\\
bcsstk09 & 1,083 & 8,677 & grid \\
block\_2000 & 2,000 & 9,912 & clusters \\
sierpinski3d & 2,050 & 6,144 & miscellaneous network\\
CA-GrQc & 4,158 & 13,422 & collaboration network \\
EVA & 4,475 & 4,652 & collaboration network \\
3elt & 4,720 & 13,722 & 3D mesh \\
us\_powergrid & 4,941 & 6,594 & miscellaneous network\\
G65 & 8,000 & 16,000 & 3D torus \\
fe\_4elt2 & 11,143 & 32,818 & 3D mesh \\
%barth5 & 15,606 & 45,878 & mesh-like \\
bcsstk31 & 32,715 & 572,914 & 3D automobile component \\
venkat50 & 62,424 & 827,671 & 3D mesh \\
ship\_003 & 121,728 & 1,827,654 & 3D ship \\
%bitcoin20w & 207,689 & 547,500 & payment network\\
%fe\_ocean & 143,437 & 409,593 & mesh-like \\
troll & 213,453 & 5,885,829 & 3D structure \\
% torso3 & 259,156 & 2,085,329 & 3D mesh \\
web-NotreDame & 325,729 & 1,469,679 & web graph\\
%Hook\_1498 & 1,498,023 & 29,709,711 & 3D steel hook \\
Flan\_1565 & 1,564,794 & 56,300,289 & 3D steel flange \\
com-Orkut & 3,072,441 & 117,185,083 & online social network\\
com-LiveJournal & 3,997,962 & 34,681,189 & online social network\\
\bottomrule
\end{tabular}
}
\end{table}

\textbf{Methods.}
We compare DRGraph with seven widely used graph layout algorithms.
We choose spring-electrical approach (FR~\cite{fruchterman1991graph}), energy-based approaches (KK~\cite{kamada1989algorithm} and Stress Majorization~\cite{gansner2004graph}), multi-level methods ($\text{FM}^3$~\cite{hachul2004drawing} and SFDP \cite{hu2005efficient}), and landmark-based algorithm (PMDS~\cite{brandes2006eigensolver}), because they are representatives of well-established approaches.
tsNET \cite{kruiger2017graph} is the state-of-the-art graph layout approache that best preserves neighborhood information.
% The unique difference lies in that tsNET* assigns initial values by PMDS. 
The implementations of FR, KK, Stress Majorization (S.M.), $\text{FM}^3$, and PMDS are gathered from OGDF-2018-03-28 ~\cite{chimani2013open}.
tsNET~\cite{kruiger2017graph} is provided by the authors.
We accelerate tsNET with a GPU-based t-SNE implementations \cite{chan2018t}.
We employ the SFDP implementation of the GraphViz library.
We repeat the experiments five times to remove the random effects.
%We repeat the experiments five times and use the mean value to remove the random effects.

\textbf{Parameters.}
After a preliminary evaluation, we set the number of negative samples to be 5,  $\gamma$=0.1, and the total number of iterations to be $400$.
DRGraph approximates the node similarity using the first-order nearest neighbors.
Due to the space limit, we discuss parameter sensitivity in the supplementary material.
%For PMDS, we set the pivot number to be the square root of the node size.
We use the pre-set parameters for other methods. 

\subsection{Evaluation Metrics}
We employ neighborhood preservation (NP), stress, crosslessness and minimum angle metrics to evaluate the graph layout quantitatively.

\textbf{Neighborhood preservation.} 
NP is defined as the Jaccard similarity coefficient between the graph space and the layout space:
\begin{equation}
NP=\frac{1}{|V|}\sum_i\frac{|NNG(v_i,k^{eval})\cap NNL(y_i,k_i^{eval})|}{|NNG(v_i,k^{eval})\cup NNL(y_i,k_i^{eval})|},
\end{equation}
where $NNG(v_i,k^{eval})$ denotes the $k^{eval}$-order nearest neighborhoods of node $v_i$ in the graph space and $NNL(y_i,k_i^{eval})$ is the $k_i^{eval}$-nearest neighbors ($k_i=|NNG(v_i,k^{eval})|$) of node $y_i$ in the layout space.
We evaluate the accuracy of neighborhood preservation with $k^{eval}=2$~\cite{kruiger2017graph}.

\textbf{Stress.} The normalized stress measures how the graph layout fits theoretical distances.
For fair comparisons, we find a scalar $\alpha$ to minimize the full stress:
\begin{equation}
    stress=\min_\alpha\frac{1}{|V|^2}\sum_{0<i,j< V}w_{ij}(\alpha\|y_i-y_j\|-SPD(v_i,v_j))^2.
\end{equation}
We use the conventional weighting factor of $w_{ij}=1/SPD(v_i,v_j)^2$.

\textbf{Crosslessness.} The crosslessness aesthetic metric \cite{purchase2002metrics} encourages graph layout methods to minimize the number of edge crosses. Inspired by it, we define the crosslessness as:
\begin{equation}
crosslessness=\left\{
\begin{aligned}
&1-\sqrt{\frac{c}{c_{max}}},   & \text{if }c_{max}>0 \\
&1,   & \text{otherwise}
\end{aligned}
\right.
\end{equation}
\begin{equation}
c_{max}=\frac{|E|(|E|-1)}{2}-\frac{1}{2}\sum_{v\in V}\text{degree}(v)(\text{degree}(v)-1),
\end{equation}
where $c$ is the number of crossings and $c_{max}$ is the approximated upper bound on the number of edge crosses.

\textbf{Minimum angle.} The minimum angle metric quantifies the average deviation of the actual minimum angle from the ideal angle \cite{purchase2002metrics}:
\begin{equation}
min\_angle=1-\frac{1}{|V|}\sum_{v\in V}|\frac{\theta(v)-\theta_{min}(v)}{\theta(v)}|, \theta(v)=\frac{360}{degree(v)},
\end{equation}
where $\theta_{min}(v)$ is the actual minimum angle at node $v$.

% Therefore, we adopt the \textit{neighborhood stress} used by Meyerhenke et al.~\cite{7889042}: 
% $$stress = \frac{1}{|E|}\sum_{(v_i,v_j)\in E}GD_{ij}^{-2}(GD_{ij} - \alpha LD_{ij})^2$$
% The \textit{neighborhood stress} is computed by finding the scalar $\alpha$ which minimizes the stress.

\subsection{Selection of Parameters}

\begin{figure}[!h]
  \centering
  \includegraphics[width=0.99\columnwidth]{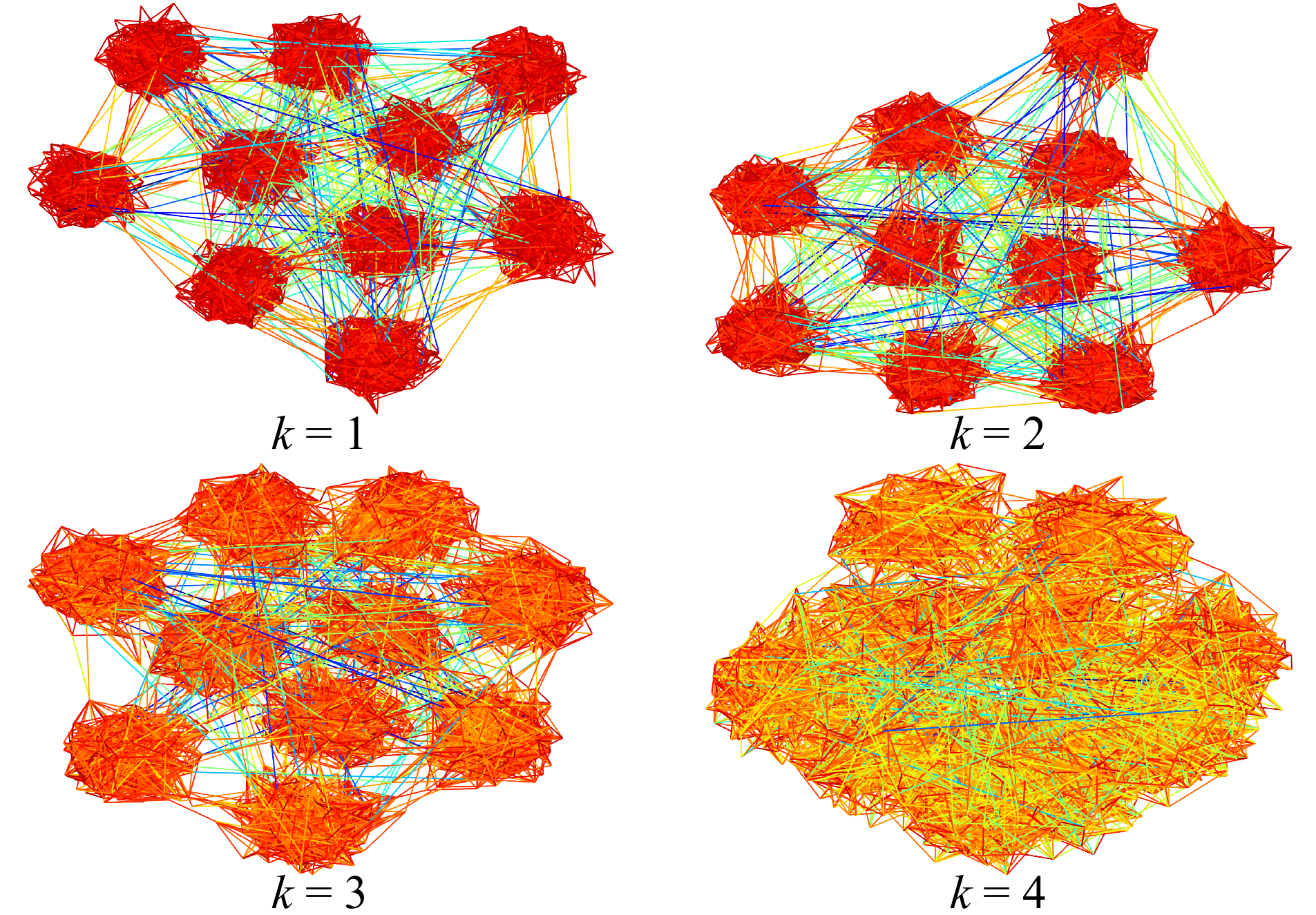}
  \caption{The effect of the size of $k$-order nearest neighbors.}
  \label{figure:k}
\end{figure}

\textbf{The size of $k$-order nearest neighbors.}
Higher-order nearest neighbors contain many dissimilar nodes, which are treated as positive nodes by the negative sampling technique.
Thus, as shown in Figure \ref{figure:k}, DRGraph places dissimilar nodes close to one another resulting in a low graph layout quality.
In addition, higher-order nearest neighbors cost large memory consumptions for keeping graph distances.
Therefore, we choose first-order nearest neighbors, which is sufficient to provide locality properties, accelerate the computation of the node similarity, and meanwhile reduce the memory requirement.	

\textbf{The number of negative samples $\text{M}$.}
When the number of negative samples becomes large adequately, the graph layout quality becomes stable.
However, the computation complexity of the optimization process is linear with $\text{M}$.
Therefore we choose $\text{M}$ to be 5 to maintain the balance between quality and efficiency.

\textbf{The weight of negative samples $\gamma$.} 
$\gamma$ controls the value of repulsive forces of the gradient. 
A small value of $\gamma$ generates small repulsive forces, whereas natural clusters in the graph data tend to form groups.
Thus, it is easier for similar nodes to move to one another in the early optimization process.
DRGraph employs the early exaggeration technique to find a better solution.
We set $\gamma$ to be 0.01 for the coarsest graph to decrease the repulsive forces and form separated clusters.
Then, we increase $\gamma$ for other coarse graphs and place nodes evenly for pleasing visualization.
NP, crosslessness, and minimum angle metrics slightly drop when $\gamma$ is small.
A large $\gamma$ leads to a bad stress quality.
We choose a medium value of 0.1 for finer graphs.

\textbf{The iteration number T.}
The layout quality becomes stable when the iteration number $\text{T}$ is large adequately.
We choose the iteration number $\text{T}=400$ to generate comparable results.

\begin{table}[!t]\setlength{\tabcolsep}{4pt}
\caption{
Time consumptions (second) of different graph layout algorithms. 
}
\label{table:runningtime}
\scalebox{0.8}{
\begin{tabular}{@{}lllllllll@{}}
\toprule
Datasets & FR & KK & S.M. & $\text{FM}^3$ & SFDP & PMDS & tsNET & DRGraph\\ \midrule
dwt\_72 & 0.006 & 0.003 & 0.011 & 0.018 & 0.006 & \textbf{0.001} & 1.727 & 0.007 \\
lesmis & 0.008 & 0.003 & 0.010 & 0.009 & 0.009 & \textbf{0.001} & 1.776 & 0.007 \\
can\_96 & 0.010 & 0.005 & 0.016 & 0.012 & 0.010 & \textbf{0.001} & 1.767 & 0.009 \\
rajat11 & 0.016 & 0.010 & 0.025 & 0.021 & 0.015 & \textbf{0.002} & 1.772 & 0.015 \\
jazz & 0.053 & 0.026 & 0.064 & 0.063 & 0.038 & \textbf{0.010} & 1.800 & 0.019 \\
visbrazil & 0.035 & 0.026 & 0.077 & 0.046 & 0.052 & \textbf{0.008} & 1.753 & 0.026 \\
grid17 & 0.059 & 0.046 & 0.121 & 0.058 & 0.030 & \textbf{0.014} & 1.818 & 0.085 \\
mesh3e1 & 0.062 & 0.046 & 0.127 & 0.065 & 0.031 & \textbf{0.012} & 1.793 & 0.079 \\
netscience & 0.100 & 0.081 & 0.214 & 0.092 & 0.051 & \textbf{0.015} & 1.844 & 0.039 \\
dwt\_419 & 0.128 & 0.102 & 0.262 & 0.101 & 0.054 & \textbf{0.018} & 1.793 & 0.043 \\
price\_1000 & 0.627 & 0.665 & 1.480 & 0.204 & 0.283 & \textbf{0.040} & 1.941 & 0.121 \\
dwt\_1005 & 0.663 & 0.686 & 1.500 & 0.141 & 0.152 & \textbf{0.048} & 1.955 & 0.114 \\
cage8 & 0.687 & 0.728 & 1.522 & 0.161 & 0.162 & \textbf{0.058} & 1.992 & 0.111 \\
bcsstk09 & 0.836 & 0.850 & 1.718 & 0.166 & 0.175 & \textbf{0.061} & 2.013 & 0.282 \\
block\_2000 & 2.603 & 3.050 & 6.174 & 0.390 & 0.382 & \textbf{0.131} & 2.217 & 0.224 \\
sierpinski3d & 2.682 & 3.007 & 6.309 & 0.317 & 0.312 & \textbf{0.092 }& 2.078 & 0.235 \\
CA-GrQc & 10.81 & 13.44 & 26.24 & 1.015 & 0.953 & \textbf{0.222} & 3.440 & 0.535 \\
EVA & 12.45 & 15.11 & 30.01 & 0.781 & 1.330 & \textbf{0.191} & 3.218 & 0.493 \\
3elt & 14.03 & 16.89 & 33.74 & 0.849 & 0.858 & \textbf{0.235} & 2.843 & 0.563 \\
us\_powergrid & 15.07 & 18.21 & 36.54 & 1.168 & 0.979 & \textbf{0.237} & 3.071 & 0.702 \\
G65 & 40.01 & 53.59 & 92.75 & 1.477 & 1.427 & \textbf{0.374} & 3.481 & 1.079 \\
fe\_4elt2 & 77.88 & 136.6 & 183.3 & 1.975 & 2.243 & \textbf{0.563} & 5.533 & 1.482 \\
bcsstk31 & 792.6 & 3916 & 2358 & 8.171 & 13.08 & \textbf{5.446} & (-) & 8.928 \\
venkat50 & 2395 & (-) & 6848 & 14.92 & 22.55 & \textbf{8.213} & (-) & 17.29 \\
ship\_003 & 9023 & (-) & (-) & 36.86 & 56.17 & \textbf{22.86} & (-) & 36.79 \\
troll & (-) & (-) & (-) & 63.14 & 122.6 & \textbf{53.22} & (-) & 66.35 \\
% torso3 & (-) & (-) & (-) & 67.00 & 139.2 & 35.85 & (-) & 83.57 \\
Web-NotreDame & (-) & (-) & (-) & 107.9 & 300.1 & \textbf{35.74} & (-) & 111.9 \\
Flan\_1565 & (-) & (-) & (-) & 623.8 & 1395 & \textbf{490.7} & (-) & 823.5 \\
com-Orkut & (-) & (-) & (-) & (-) & (-) & 4444 & (-) & \textbf{1994} \\
com-LiveJournal & (-) & (-) & (-) & 3066 & 7269 & \textbf{1644} & (-) & 2943
\\ \bottomrule
\end{tabular}
}
\end{table}

\begin{figure}[!t]
  \centering
  \includegraphics[width=0.99\columnwidth]{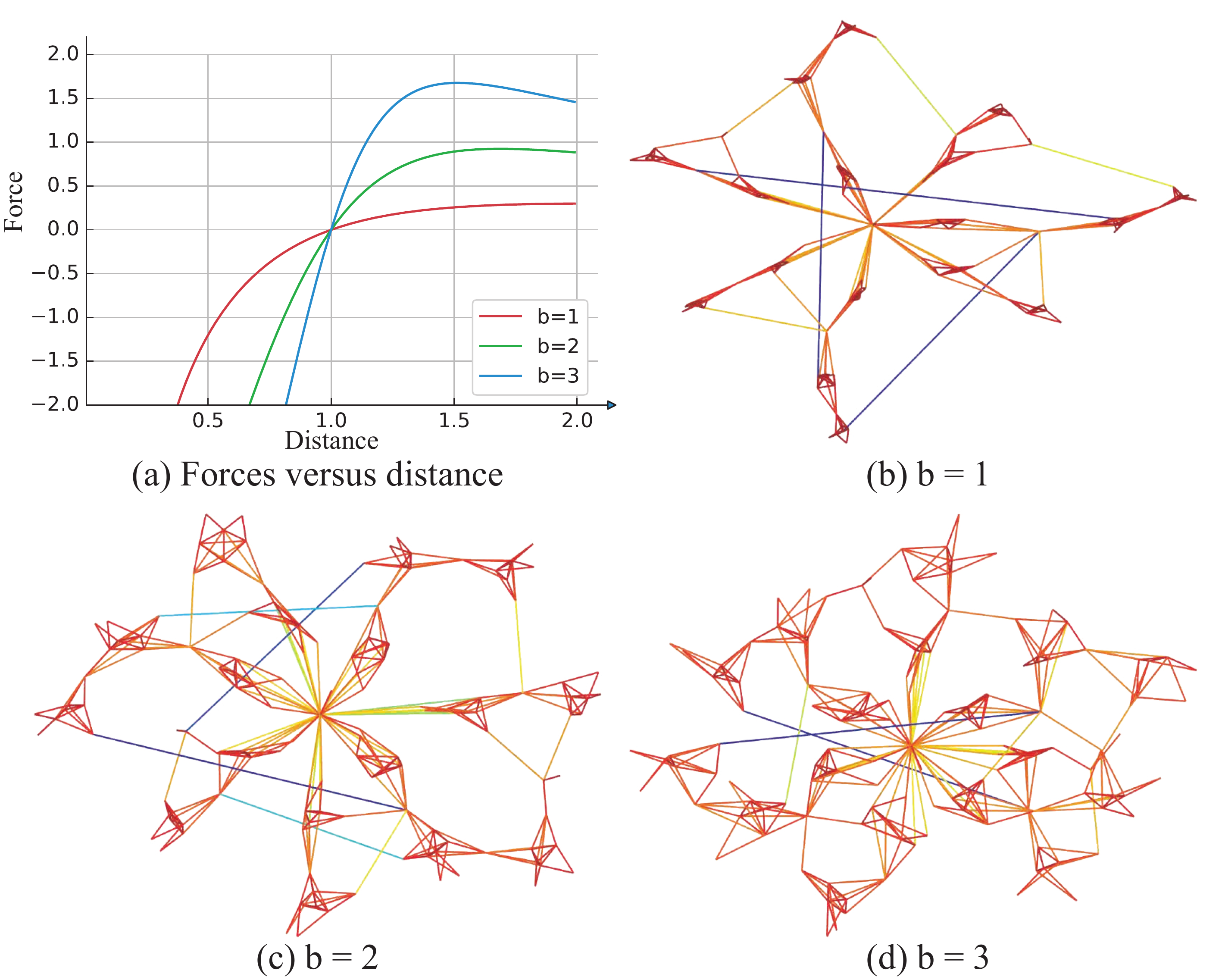}
  \caption{The effect of the $b$ parameter.}
  \label{figure:b}
\end{figure}

\textbf{The effect of $\textbf{b}$.}
Figure \ref{figure:b} (a) illustrates the sum of the attractive and the repulsive forces (\ie, the gradient defined in Eq. \ref{eq:gradient}) with respect to $b$.
$b$ controls the value of the sum force without altering the ideal distance between nodes.
Generally, a small $b$ value (\eg, $b = 1$) tends to place nodes close to others and generates localized clusters (Figure \ref{figure:b} (b)).
A large $b$ value (\eg, $b = 3$) forces all edge lengths to be ideal but obfuscates the global structure (Figure \ref{figure:b} (d)).
For 3D meshes (\eg, G65) and large social networks (\eg, com-Orkut), preserving all edge lengths of a manifold into the 2D space is intractable.
%that are obtained by discretizing 3D shapes with finite elements
Therefore, we set $b=1$ to preserve the neighborhood identity for these graphs.
We choose $b=3$ when the input graph is a grid graph (\eg, grid17), in which all edges have the same length.
For other graphs, we choose $b=2$, which works well in preserving local and global structures.

\begin{table}[!t]\setlength{\tabcolsep}{4pt}
\caption{Comparisons of the memory consumption (MB).}
\label{table:space}
\scalebox{0.8}{
\begin{tabular}{@{}lllllllll@{}}
\toprule
Datasets & FR & KK & S.M. & $\text{FM}^3$ & SFDP & PMDS & tsNET & DRGraph\\ \midrule
dwt\_72 & \textbf{4} & \textbf{4} & 5 & 5 & 6 & \textbf{4} & 284 & 5 \\
lesmis & \textbf{4} & \textbf{4} & 5 & 5 & 7 & \textbf{4} & 277 & 5 \\
can\_96 & \textbf{4} & \textbf{4} & 5 & 5 & 7 & \textbf{4} & 278 & 5 \\
rajat11 & \textbf{4} & \textbf{4} & 5 & 5 & 7 & \textbf{4} & 278 & 5 \\
jazz & \textbf{4} & 5 & 7 & 7 & 9 & 5 & 284 & 5 \\
visbrazil & \textbf{4} & 5 & 6 & 5 & 7 & \textbf{4} & 280 & 5 \\
grid17 & \textbf{4} & 6 & 7 & 6 & 7 & {5} & 282 & {5} \\
mesh3e1 & \textbf{4} & 6 & 7 & 6 & 7 & {5} & 283 & {5} \\
netscience & \textbf{4} & 7 & 8 & 6 & 7 & {5} & 284 & {5} \\
dwt\_419 & \textbf{4} & 7 & 9 & 6 & 8 & {5} & 285 & {5} \\
price\_1000 & \textbf{4} & 20 & 21 & 6 & 9 & 6 & 322 & {5} \\
dwt\_1005 & \textbf{5} & 20 & 22 & 8 & 10 & 7 & 320 & 6 \\
cage8 & \textbf{5} & 21 & 23 & 10 & 11 & 7 & 322 & 6 \\
bcsstk09 & \textbf{5} & 36 & 43 & 13 & 14 & 8 & 329 & 6 \\
block\_2000 & \textbf{6} & 69 & 73 & 15 & 16 & 10 & 446 & 7 \\
sierpinski3d & \textbf{5} & 110 & 137 & 13 & 14 & 9 & 454 & 7 \\
CA-GrQc & \textbf{7} & 416 & 534 & 23 & 22 & 15 & 889 & 9 \\
EVA & \textbf{5} & 456 & 570 & 16 & 16 & 14 & 994 & 8 \\
3elt & \textbf{7} & 491 & 605 & 22 & 23 & 16 & 1077 & 9 \\
us\_powergrid & \textbf{6} & 521 & 629 & 18 & 19 & 15 & 1160 & 9 \\
G65 & \textbf{8} & 1009 & 1018 & 25 & 30 & 23 & 3099 & 11 \\
fe\_4elt2 & \textbf{12} & 2399 & 2818 & 48 & 46 & 33 & 4057 & 15 \\
bcsstk31 & 130 & 27730 & 35933 & 539 & 505 & 193 & (-) & \textbf{78} \\
venkat50 & 173 & (-) & 62255 & 668 & 701 & 283 & (-) & \textbf{114} \\
ship\_003 & 368 & (-) & (-) & 1626 & 1545 & 580 & (-) & \textbf{278} \\
troll & (-) & (-) & (-) & 4712 & 4557 & 1550 & (-) & \textbf{670} \\
% torso3 & (-) & (-) & (-) & 1837 & 1946 & 895 & (-) & 383 \\
web-NotreDame & (-) & (-) & (-) & 1449 & 1522 & 950 & (-) & \textbf{343} \\
Flan\_1565 & (-) & (-) & (-) & 44408 & 43651 & 13437 & (-) & \textbf{6404} \\
com-Orkut & (-) & (-) & (-) & (-) & (-) & 27465 & (-) & \textbf{18411} \\
com-LiveJournal & (-) & (-) & (-) & 42443 & 31686 & 15220 & (-) & \textbf{7450}
\\ \bottomrule
\end{tabular}
}
\end{table}

\begin{figure}[!t]
  \centering
  \includegraphics[width=0.99\columnwidth]{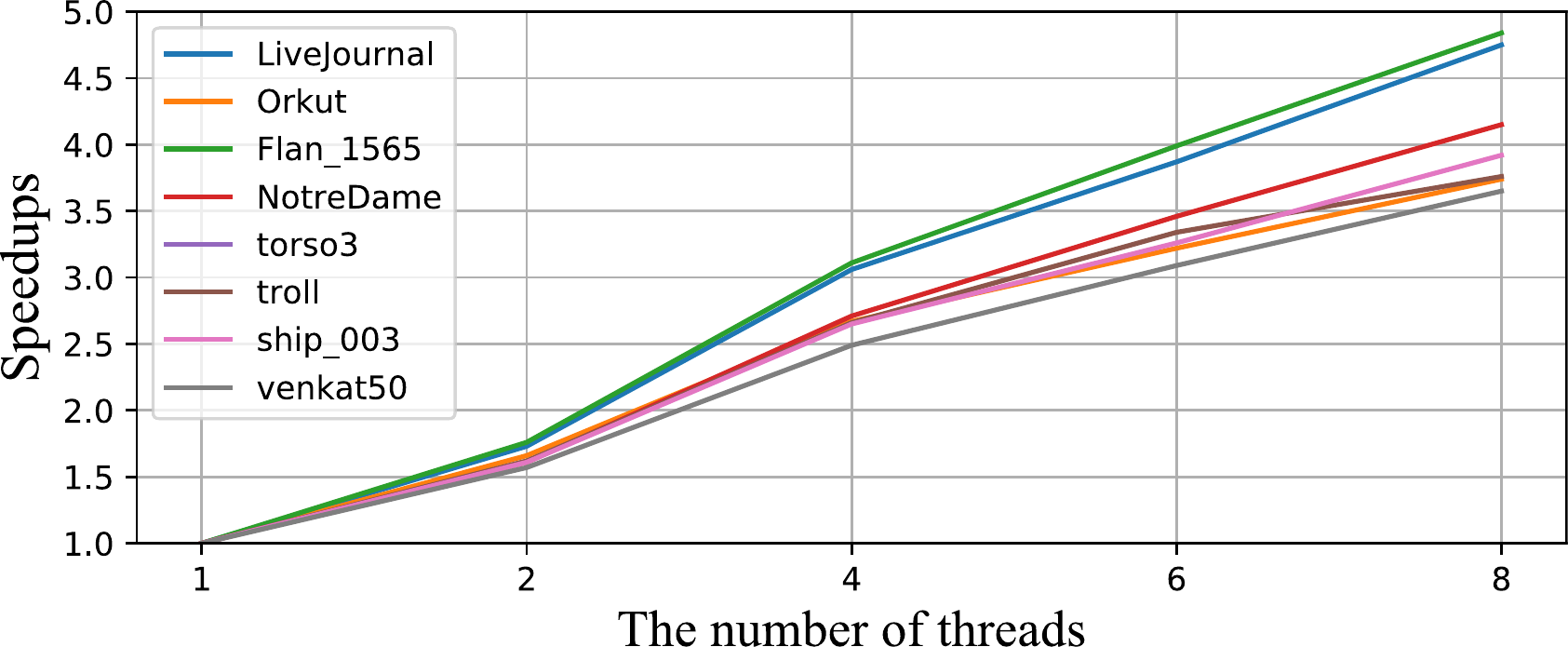}
  \caption{The parallel speedups of our DRGraph.}
  \label{figure:parallelperformance}
\end{figure}

\begin{table}[!t]
\setlength{\tabcolsep}{4pt}
\caption{Comparisons of the accuracy of neighborhood preservation.
}
\label{table:layoutquality1}
\scalebox{0.8}{
\begin{tabular}{@{}lllllllll@{}}
\toprule
Datasets & FR & KK & S.M. & $\text{FM}^3$ & SFDP & PMDS & tsNET & DRGraph\\ \midrule
dwt\_72 & .4601 & .8075 & \textbf{.9278} & .7334 & .8737 & .6714 & .7891 & .8822 \\
lesmis & .6812 & .6712 & .6902 & .6902 & .7170 & \textbf{.7548} & .7096 & .6558 \\
can\_96 & .5183 & .5517 & .5527 & .5383 & .5304 & .4733 & .6275 & \textbf{.6522} \\
rajat11 & .5205 & .6146 & .6231 & .6123 & .6113 & .6102 & .7017 & \textbf{.7078} \\
jazz & .8300 & .8059 & .8294 & .8392 & .8395 & \textbf{.8438} & .8004 & .7764 \\
visbrazil & .3645 & .3830 & .3860 & .4214 & .4619 & .3794 & \textbf{.5447} & .4885 \\
grid17 & .3140 & .9999 & \textbf{1.000} & .8261 & .7676 & .6870 & .7647 & .8502 \\
mesh3e1 & .3769 & .9280 & \textbf{1.000} & .9892 & .8340 & .9678 & .8988 & .9956 \\
netscience & .4568 & .4844 & .5013 & .5677 & .6030 & .4444 & \textbf{.7130} & .6657 \\
dwt\_419 & .3823 & .6883 & \textbf{.7589} & .7265 & .7218 & .6928 & .7111 & .7531 \\
price\_1000 & .3001 & .1814 & .2057 & .4331 & .4785 & .3697 & \textbf{.5882} & .5503 \\
dwt\_1005 & .2748 & .5244 & .5617 & .5354 & .4990 & .4661 & \textbf{.6201} & .5936 \\
cage8 & .2089 & .1899 & .1988 & .2044 & .2210 & .2063 & \textbf{.4240} & .2919 \\
bcsstk09 & .4655 & .9575 & \textbf{.9676} & .9015 & .8260 & .6957 & .8775 & .8882 \\
block\_2000 & .2738 & .1586 & .1597 & .2516 & .2743 & .1626 & \textbf{.3635} & .3032 \\
sierpinski3d & .1886 & .5001 & .5394 & .5198 & .5100 & .2032 & .5535 & \textbf{.5702} \\
CA-GrQc & .1186 & .0171 & .0924 & .1287 & .1481 & .1472 & \textbf{.4418} & .1722 \\
EVA & .6408 & .2028 & .4211 & .6342 & .6627 & .7037 & \textbf{.7691} & .7148 \\
3elt & .0679 & .4426 & .5121 & .6353 & .6306 & .3595 & .5824 & \textbf{.6431} \\
us\_powergrid & .0593 & .1327 & .1450 & .3092 & .3741 & .1835 & .4014 & \textbf{.4583} \\
G65 & .0241 & .2154 & .2478 & .2250 & .2273 & .1915 & \textbf{.3261} & .2594 \\
fe\_4elt2 & .0348 & .3267 & .4304 & .4840 & .4279 & .2468 & .4656 & \textbf{.5885} \\
bcsstk31 & .0989 & .2124 & .3254 & .3394 & .3656 & .2264 & (-) & \textbf{.3783} \\
venkat50 & .0552 & (-) & .4300 & .6235 & .5839 & .3178 & (-) & \textbf{.6418} \\
ship\_003 & .0501 & (-) & (-) & .1350 & .1562 & .1380 & (-) & \textbf{.1958} \\
troll & (-) & (-) & (-) & .1962 & .2121 & .1072 & (-) & \textbf{.2529} \\
web-NotreDame & (-) & (-) & (-) & \textbf{.5018} & .3771 & .3894 & (-) & .4651 \\
Flan\_1565 & (-) & (-) & (-) & .1853 & .1671 & .0934 & (-) & \textbf{.2046}
\\ \bottomrule
\end{tabular}
}
\end{table}

\begin{table}[!t]
\setlength{\tabcolsep}{4pt}
%\centering
\caption{Comparisons of the stress metric.
%The number represents the average accuracy of 10 repetitions.
}
\label{table:layoutquality2}
\scalebox{0.8}{
\begin{tabular}{@{}lllllllll@{}}
\toprule
Datasets & FR & KK & S.M. & $\text{FM}^3$ & SFDP & PMDS & tsNET & DRGraph\\ \midrule
dwt\_72 & .1564 & .0452 & \textbf{.0284} & .0673 & .0471 & .0727 & .0491 & .0435 \\
lesmis & .1294 & .0862 & \textbf{.0814} & .1000 & .1223 & .1527 & .1031 & .1265 \\
can\_96 & .1018 & \textbf{.0711} & .0732 & .0735 & .0736 & .0862 & .0841 & .0864 \\
rajat11 & .1222 & .0717 & \textbf{.0628} & .0814 & .0904 & .1159 & .0952 & .0954 \\
jazz & .1469 & .1153 & \textbf{.1014} & .1201 & .1457 & .1517 & .1329 & .1366 \\
visbrazil & .1594 & .0635 & \textbf{.0602} & .0964 & .0844 & .1557 & .0937 & .0849 \\
grid17 & .1880 & .0137 & \textbf{.0136} & .0157 & .0192 & .0270 & .0211 & .0149 \\
mesh3e1 & .1681 & .0161 & \textbf{.0025} & .0046 & .0151 & .0044 & .0135 & .0040 \\
netscience & .1620 & .0622 & \textbf{.0564} & .0940 & .1103 & .1032 & .1124 & .0948 \\
dwt\_419 & .1861 & .0372 & \textbf{.0156} & .0256 & .0347 & .0242 & .1283 & .0187 \\
price\_1000 & .1880 & .1073 & \textbf{.0925} & .1541 & .1267 & .2329 & .1680 & .1488 \\
dwt\_1005 & .2156 & .0535 & \textbf{.0212} & .0266 & .0296 & .0292 & .1243 & .0402 \\
cage8 & .1568 & .1199 & \textbf{.1181} & .1288 & .1393 & .1414 & .1855 & .1475 \\
bcsstk09 & .1365 & .0192 & \textbf{.0153} & .0173 & .0231 & .0384 & .0207 & .0167 \\
block\_2000 & .1685 & .1408 & \textbf{.1398} & .1544 & .1632 & .1763 & .1941 & .1769 \\
sierpinski3d & .3173 & .0749 & \textbf{.0626} & .0725 & .0794 & .1040 & .0991 & .0757 \\
CA-GrQc & .1662 & .1935 & \textbf{.1227} & .1398 & .1460 & .1809 & .1891 & .1872 \\
EVA & .1959 & .1725 & \textbf{.0972} & .1228 & .1288 & .2400 & .1507 & .1496 \\
3elt & .3382 & .0677 & \textbf{.0379} & .0627 & .0664 & .0562 & .1323 & .0626 \\
us\_powergrid & .2752 & .0709 & \textbf{.0573} & .1163 & .0968 & .0913 & .1901 & .1144 \\
G65 & .4161 & .1540 & \textbf{.1094} & .1147 & .1125 & .1356 & .1424 & .1110 \\
fe\_4elt2 & .4678 & .1708 & \textbf{.0455} & .0514 & .0553 & .0688 & .0707 & .0741 \\
bcsstk31 & .2862 & .1525 & .0649 & \textbf{.0457} & .0645 & .0539 & (-) & .0649 \\
venkat50 & .3261 & (-) & .0736 & .0770 & .0644 & .1205 & (-) & \textbf{.0589} \\
ship\_003 & .2566 & (-) & (-) & .0598 & \textbf{.0374} & .0469 & (-) & .0425 \\
troll & (-) & (-) & (-) & \textbf{.0656} & .1151 & .0800 & (-) & .0961 \\
web-NotreDame & (-) & (-) & (-) & \textbf{.1212} & .1385 & .1247 & (-) & .1338 \\
Flan\_1565 & (-) & (-) & (-) & \textbf{.0626} & .0886 & .0942 & (-) & .0922 \\
com-Orkut & (-) & (-) & (-) & (-) & (-) & \textbf{.2005} & (-) & .2055 \\
com-LiveJournal & (-) & (-) & (-) & \textbf{.1470} & .1602 & .1871 & (-) & .2018
\\ \bottomrule
\end{tabular}
}
\end{table}

\subsection{Performance}
\label{section:runningtime}
\textbf{Running time.}
Table \ref{table:runningtime} reports the running time of graph visualization process.
For all approaches, the running time only includes the layout time without data process steps.
We employ the single-thread version of DRGraph for a fair comparison.
Unfilled items indicate the incapability of the corresponding algorithm caused by the huge memory consumption or computational cost.
For small datasets, most graph layout methods perform comparably to each other.
Especially, the single-thread version of DRGraph is faster than GPU-accelerated tsNET.
For large datasets, $\text{FM}^3$, DRGraph, and PMDS are much more efficient than others.
PMDS is the fastest method due to the number of pivots used.
Our DRGraph runs faster than PMDS on the com-Orkut dataset.
The performance of PMDS is severely affected by the number of edges.
For the Flan\_1565 and com-LiveJournal datasets, DRGraph, $\text{FM}^3$, SFDP and PMDS are comparable in terms of the running time.
$\text{FM}^3$ and SFDP fail to visualize the com-Orkut dataset due to their huge memory consumptions.
Though multilevel-based graph layout method $\text{FM}^3$ achieves comparable performance on large-scale datasets, DRGraph requires less memory consumption and generates results with better NP than $\text{FM}^3$ (see Section \ref{section:layoutquality}).

The parallel implementation of optimization enables further acceleration on a multi-core platform. 
Figure \ref{figure:parallelperformance} plots the speedups in terms of the number of threads.
Generally, the speedups increase with data sizes and the number of threads.
The largest overall speedup (4.84$\times$) is obtained by eight threads on the Flan\_1565 dataset.
DRGraph reduces the running time of Flan\_1565 from 823.5 seconds to 171.1 seconds using eight threads.
For the com-Orkut dataset, DRGraph spends much time on graph coarsening, leading to a slightly small speedup (3.74$\times$).

\textbf{Memory consumption.}
Table \ref{table:space} compares the memory usages.
The memory usage denotes the maximum usage of the process during its lifetime.
Energy models such as KK and S.M. are huge consumers of memory because they require quadratic memory complexity to store pairwise graph distances.
DRGraph only consumes 7 GB memory to visualize Flan\_1565 with 1,564,794 nodes and 56,300,289 edges.
Contrarily, $\text{FM}^3$ requires approximately 44 GB.
Fundamentally, DRGraph achieves a linear complexity of memory consumption $O(|E|+|V|)$ and scales up to large graphs with millions of nodes.

%Figure \ref{figure:performance} shows the running time and the memory consumption for dwt\_419, dwt\_1005, block\_2000, 3elt, G65, fe\_4elt, bcsstk31, ship\_003, troll and Flan\_1565 using DRGraph, $\text{FM}^3$, and PMDS. 
%They are the best algorithms among nine methods in terms of both the time and memory consumptions.
%Each dot represents the performance of a graph layout method for a dataset.
%The dot color encodes the type of the employed method.
%For each layout method, we connect dots with a line in the order of the data size.
%We represent the node size as a label for each dot.
%Here, we compare the performance between DRGraph, $\text{FM}^3$ and PMDS, because they can handle large-scale datasets.
%As the number of nodes increases, the graph between DRGraph and other layout methods enlarges in both running time and memory consumption.
%
%\begin{figure}[!ht]
%  \centering
%  \includegraphics[width=0.99\columnwidth]{./picture/performance.eps}
%  \caption{
%  The (time consumption, memory consumption) dot plot for 10 datasets (dwt\_419, dwt\_1005, block\_2000, 3elt, G65, fe\_4elt, bcsstk31, ship\_003, troll and Flan\_1565) with three methods.
%  Because the value range is very large, both the ranges along the $x$ axis and $y$ axis are logarithmically scaled.
%  The dot color represents the type of the employed method.
%  For each layout method, we connect dots with a line in the order of the size of the dataset.
%  DRGraph is increasingly superior to other graph layout methods when the data size increases. 
%  }
%  \label{figure:performance}
%\end{figure}

\begin{table}[!t]
\setlength{\tabcolsep}{4pt}
%\centering
\caption{Comparisons of the crosslessness metric.
%The number represents the average accuracy of 10 repetitions.
}
\label{table:layoutquality3}
\scalebox{0.8}{
\begin{tabular}{@{}llllllllll@{}}
\toprule
Datasets & FR & KK & S.M. & $\text{FM}^3$ & SFDP & PMDS & tsNET & DRGraph\\ \midrule
dwt\_72 & .9559 & .9606 & .9389 & .9695 & \textbf{1.000} & .9614 & .9727 & .9817 \\
lesmis & .7756 & .7631 & .7429 & .7784 & .8306 & .7527 & .8026 & \textbf{.8323} \\
can\_96 & .8841 & .8922 & .8974 & .8946 & .9257 & .8964 & .8899 & \textbf{.9355} \\
rajat11 & .8791 & .8749 & .8673 & .8790 & \textbf{.9335} & .8671 & .8841 & .9292 \\
jazz & .6930 & .6499 & .6905 & .6934 & \textbf{.7667} & .6919 & .6713 & \textbf{.7667} \\
visbrazil & .8188 & .8094 & .8398 & .8603 & .9219 & .8216 & .8538 & \textbf{.9270} \\
grid17 & .9510 & \textbf{1.000} & \textbf{1.000} & .9618 & \textbf{1.000} & \textbf{1.000} & \textbf{1.000} & \textbf{1.000} \\
mesh3e1 & .9071 & .9329 & .9403 & .9381 &\textbf{1.000} & .9452 & .8886 & \textbf{1.000} \\
netscience & .8599 & .8457 & .8750 & .8870 & .9517 & .8893 & .8886 & \textbf{.9532} \\
dwt\_419 & .8577 & .8820 & .8828 & .8960 & \textbf{.9572} & .8813 & .8952 & .9550 \\
price\_1000 & .8463 & .7752 & .8225 & .8984 & .9838 & .8361 & .9230 & \textbf{.9887} \\
dwt\_1005 & .8936 & .9057 & .9200 & .9212 & \textbf{.9687} & .9159 & .9167 & .9658 \\
cage8 & .8584 & .8512 & .8622 & .8577 & .8794 & .8511 & .8577 & \textbf{.8955} \\
bcsstk09 & .9386 & .9556 & .9568 & .9500 & \textbf{.9582} & .9384 & .9566 & .9566 \\
block\_2000 & .8003 & .7514 & .7534 & .7855 & .8865 & .7401 & .8180 & \textbf{.8908} \\
sierpinski3d & .9493 & .9516 & .9538 & .9566 & .9836 & .9478 & .9593 & \textbf{.9845} \\
CA-GrQc & .8643 & .7425 & .8527 & .8682 & .9203 & .8632 & .8823 & \textbf{.9262} \\
EVA & .8874 & .7087 & .8516 & .8889 & .9819 & .8869 & .9112 & \textbf{.9854} \\
3elt & .9645 & .9766 & .9824 & .9824 & \textbf{.9931} & .9804 & .9807 & .9927 \\
us\_powergrid & .9378 & .9404 & .9509 & .9622 & .9879 & .9511 & .9666 & \textbf{.9895} \\
G65 & .9701 & .9819 & .9853 & .9858 & .9878 & .9860 & .9838 & \textbf{.9888} \\
fe\_4elt2 & .8973 & .9331 & .9551 & .9601 & .9939 & .9535 & .9606 & \textbf{.9945} \\
bcsstk31 & .9718 & .9577 & .9684 & .9731 & \textbf{.9837} & .9702 & (-) & .9834 \\
venkat50 & .9654 & (-) & .9840 & .9854 & .9931 & .9836 & (-) & \textbf{.9933} \\
ship\_003 & .9735 & (-) & (-) & .9764 & .9835 & .9594 & (-) & \textbf{.9843} \\
troll & (-) & (-) & (-) & .9839 & \textbf{.9900} & .9771 & (-) & .9899 \\
web-NotreDame & (-) & (-) & (-) & .9360 & .9599 & .8738 & (-) & \textbf{.9715}
\\ \bottomrule
\end{tabular}
}
\end{table}

\begin{table}[!t]
\setlength{\tabcolsep}{4pt}
%\centering
\caption{Comparisons of the minimum angle metric.
%The number represents the average accuracy of 10 repetitions.
}
\label{table:layoutquality4}
\scalebox{0.8}{
\begin{tabular}{@{}llllllllll@{}}
\toprule
Datasets & FR & KK & S.M. & $\text{FM}^3$ & SFDP & PMDS & tsNET & DRGraph\\ \midrule
dwt\_72 & .8150 & .8474 & .8296 & .8536 & \textbf{.9434} & .8640 & .8836 & .9205 \\
lesmis & .3497 & .2817 & .3157 & .3311 & .4181 & .3320 & .2837 & \textbf{.4209} \\
can\_96 & .3118 & .1191 & .1702 & .2963 & .2116 & .1698 & .0000 & \textbf{.3286} \\
rajat11 & .0895 & .0663 & .0759 & .0812 & \textbf{.2471} & .0907 & .0277 & .2031 \\
jazz & .0522 & .0565 & .0694 & .0593 & .0666 & .0598 & .0330 & \textbf{.0814} \\
visbrazil & .5929 & .5915 & .5762 & .5808 & \textbf{.6618} & .5662 & .4612 & .6554 \\
grid17 & .3620 & .3998 & .2840 & .5803 & \textbf{.9123} & .4845 & .0616 & .8570 \\
mesh3e1 & .2776 & .3808 & .1974 & .3487 & .7008 & .2393 & .0167 & \textbf{.8131} \\
netscience & .2638 & .2426 & .2457 & .2526 & \textbf{.3883} & .2108 & .1436 & .3724 \\
dwt\_419 & .1020 & .1064 & .0757 & .0990 & \textbf{.2073} & .0635 & .0000 & .2033 \\
price\_1000 & .8363 & .8378 & .8396 & .8365 & \textbf{.8883} & .8363 & .8380 & .8584 \\
dwt\_1005 & .1204 & .1591 & .0916 & .1460 & .2577 & .1277 & .0000 & \textbf{.3382} \\
cage8 & .1212 & .1191 & .1232 & .1086 & \textbf{.1696} & .1242 & .0006 & .1460 \\
bcsstk09 & .0344 & .0405 & .0332 & .0289 & .0378 & .0183 & .0001 & \textbf{.0646} \\
block\_2000 & .0222 & .0671 & .0371 & .0165 & .1017 & .0149 & .0008 & \textbf{.1168} \\
sierpinski3d & .0888 & .1099 & .0610 & .0980 & .2325 & .0762 & .0002 & \textbf{.2387} \\
CA-GrQc & .3376 & .2888 & .3129 & .3315 & \textbf{.4173} & .3062 & .2502 & .4140 \\
EVA & .9085 & .9036 & .9083 & .9063 & \textbf{.9385} & .9051 & .9008 & .9315 \\
3elt & .2304 & .2545 & .2781 & .3779 & \textbf{.6025} & .3441 & .0000 & .5076 \\
us\_powergrid & .5806 & .5385 & .5544 & .5570 & \textbf{.7094} & .5308 & .4585 & .6926 \\
G65 & .4288 & .7557 & .4724 & .7237 & .4511 & \textbf{.6280} & .0038 & .6125 \\
fe\_4elt2 & .3338 & .2358 & .3425 & .3111 & .4574 & .2467 & .0014 & \textbf{.5059} \\
bcsstk31 & .0039 & .0033 & .0049 & .0027 & .0116 & .0013 & (-) & \textbf{.0246} \\
venkat50 & .0021 & (-) & .0100 & .0024 & .0050 & .0000 & (-) & \textbf{.0212} \\
ship\_003 & .0040 & (-) & (-) & .0013 & .0146 & .0002 & (-) & \textbf{.0241} \\
troll & (-) & (-) & (-) & .0004 & .0020 & .0000 & (-) & \textbf{.0092} \\
web-NotreDame & (-) & (-) & (-) & .5441 & .5452 & .5436 & (-) & \textbf{.5607} \\
Flan\_1565 & (-) & (-) & (-) & .0000 & .0013 & .0000 & (-) & \textbf{.0080} \\
com-Orkut & (-) & (-) & (-) & (-) & (-) & .0254 & (-) & \textbf{.0532} \\
com-LiveJournal & (-) & (-) & (-) & .2410 & .2960 & .2342 & (-) & \textbf{.3187} \\
\bottomrule
\end{tabular}
}
\end{table}

\subsection{Graph Layout Quality}
\label{section:layoutquality}
Tables \ref{table:layoutquality1}, \ref{table:layoutquality2}, \ref{table:layoutquality3}, and \ref{table:layoutquality4} compare NP, stress, crosslessness and minimum angle metrics of different graph layout algorithms.
FR and KK produce a poor layout quality on large graphs, because they easily converge to local minima and can hardly preserve the graph structure.
DRGraph and tsNET are superior to other methods in terms of NP due to the local structure preservation nature of $t$-SNE.
DRGraph performs slightly better than tsNET on graphs with regular structures (\eg, grid17 and 3elt).
DRGraph obtains a worse layout quality than tsNET on cage8 and CA-GrQc.
This is because the negative sampling technique cannot easily identify local and global structures of these irregular graphs.
The gap of NP metric between tsNET and DRGraph is small indicating that our method achieves a comparable layout quality.
In addition, DRGraph can achieve a better stress quality compared to tsNET.
The stress metric of DRGraph and $\text{FM}^3$ is better on small graphs than the results obtained by PMDS.
$\text{FM}^3$ reaches the best stress quality on large datasets.
This is not surprising because our algorithm does not optimize the stress. 
DRGraph has a better NP quality than $\text{FM}^3$ and SFDP on almost all graphs.
Moreover, DRGraph and SFDP achieve the best performance in terms of the crosslessness and the minimum angle aesthetic metrics.
Ultimately, DRGraph achieves a comparable layout quality to $\text{FM}^3$, SFDP and tsNET.
%We find that KK and majorization achieve very high neighborhood preservation accuracy on bcsstk09 dataset.
%The reason is that the object of neighborhood preservation is equivalent to the aim of stress function on a graph with regular structures.

\subsection{Visualization Results}
\label{section:visualization}
\begin{figure}[!h]
  \centering
  \includegraphics[width=0.99\columnwidth]{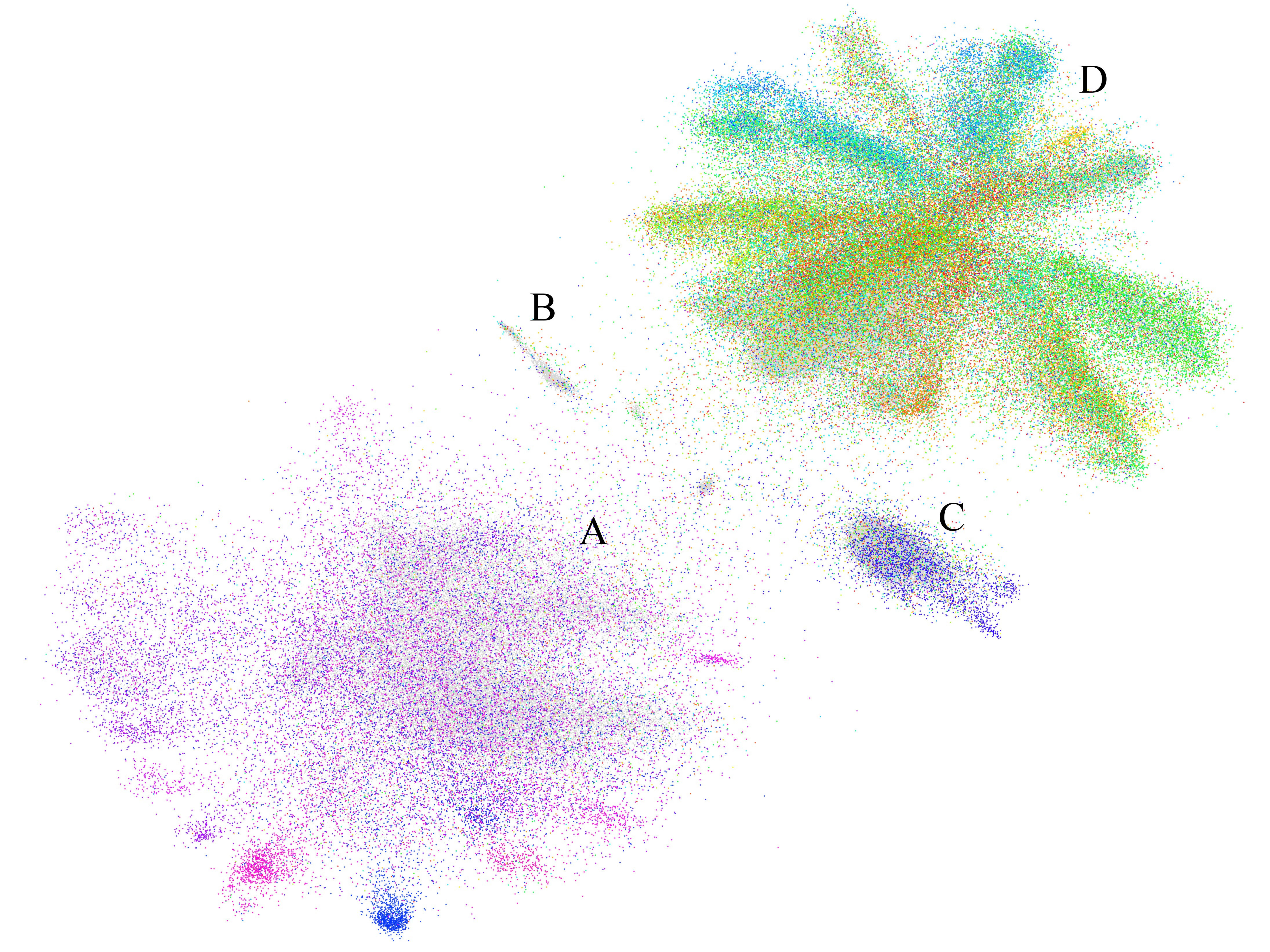}
  \caption{Visualization of the com-Orkut dataset. }
  \label{figure:casestudy}
\end{figure}

\begin{figure*}[!ht]
  \centering
  \includegraphics[width=1.99\columnwidth]{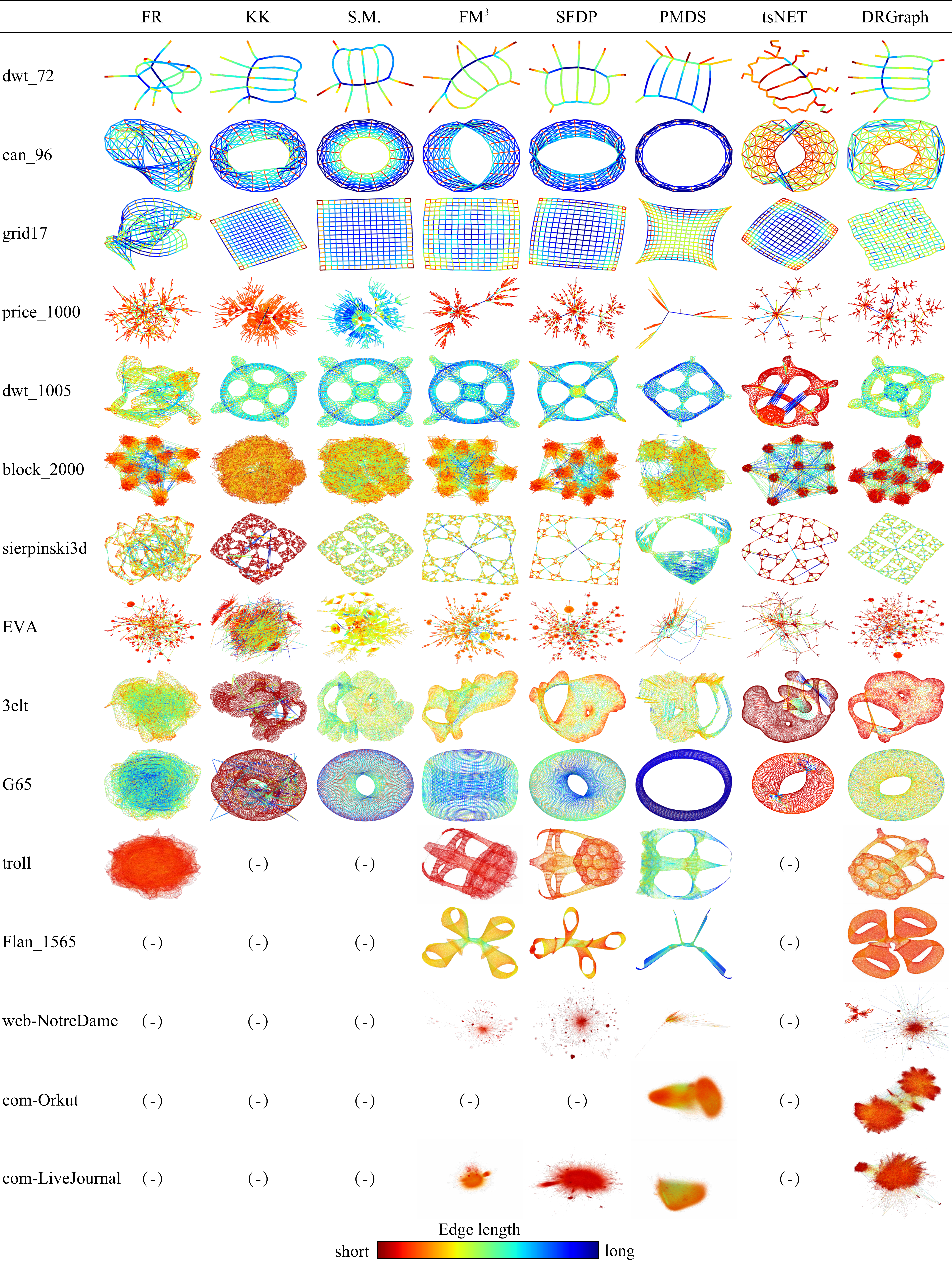}
  \caption{Visualizations of selected graph datasets using FR, KK, S.M., $\text{FM}^3$, SFDP, PMDS, tsNET and DRGraph. 
  %The edge length is encoded with the color map used in \cite{gansner2013maxent}. 
  }
  \label{figure:visualization1}
\end{figure*}

Figures \ref{figure:casestudy} and \ref{figure:visualization1} show representative graph layouts. Due to the space limit, more examples are given in supplementary material. 
We draw all graphs in Python using the Matplotlib library \cite{Hunter:2007}.
We compute the edge length from the layout space and use a red-to-green-to-blue color map to visualize the distribution of edge lengths. The shortest edge is in red, and the longest edge is in blue. Other edges are colored according to the scale.
The node-link diagram suffers from the limited screen space and possible visual clutter for visualizing large-scale datasets.
Thus, we randomly sample a subset of the edge set to reduce the visual clutter for graphs with more than 600,000 edges.
Generally, DRGraph achieves visually readable layouts (see Figure \ref{figure:visualization1}).
Force-directed methods, such as FR and KK cannot generate layouts clearly for large datasets (\eg, G65 and troll).
We can see that tsNET and DRGraph achieve aesthetically pleasing results with clear structures.
$\text{FM}^3$, SFDP, PMDS, and DRGraph usually produce better layouts than other methods on large datasets because they employ a multi-level or landmark approach to compute a better initial layout.
Results of DRGraph exhibit clearer clustering structures compared with those by $\text{FM}^3$, SFDP, and PMDS on large social networks (e.g., com-Orkut).
In Figure \ref{figure:casestudy}, we visualize users of the com-Orkut dataset by leveraging DRGraph.
Different colors encode different ground-truth communities.
We filter communities that have less than 800 users.
Roughly speaking, there are four visible groups.
Users from the same community tend to form tight clusters around the group center.
However, communities located in group A are closely connected to other users without labels.
Therefore, communities in the center of group A are visually indistinguishable from each other.

%For 3D meshes (\eg, 3elt and troll), DRGraph shows a better overall structure than $\text{FM}^3$.
%The reason is that DRGraph tends to flatten high-dimensional structures into the 2D layout space to preserve neighborhoods.

\section{Discussions}\label{sec:discussions}
\textbf{Scalability.}
Methods such as KK, Stress Majorization (S.M.), and tsNET require $O(|V|^2)$ time to compute all pairwise Euclidean distances and $O(|V|^2)$ memory to store distances.
They are not applicable to large-scale data.
Contrarily, DRGraph achieves a linear complexity for the computation and memory consumption, and can be applied to datasets with millions of nodes.

\textbf{Robustness.}
Many graph layout methods are only appropriate for limited types of graphs.
DRGraph generates satisfying graph layouts on almost all datasets with appropriate parameters.
We analyze the performance of DRGraph in planar, hierarchical, social, and tree-like graphs.
DRGraph can easily achieve good results with a similar configuration of parameters without a significant parameter modification. 
%It is also verified that the performance of DRGraph is very stable with different parameter settings.

\textbf{Generalizability.}
Graph layout methods can be unified as an optimization problem\cite{yang2014optimization}.
The difference between graph layout methods lies in the selection of node similarity, layout proximity, and distribution distance functions.
Various configurations yield distinctive graph layout approaches.
For instance, DRGraph chooses the same functions used in tsNET to keep the neighbor-preserving nature of $t$-SNE.
DRGraph distinguishes itself from others in that it utilizes a sparse distance matrix, the negative sampling technique, and a multi-level approach to accelerate the computation.
%DRGraph compares favorably with tsNET in terms of both performance and quality. 
In addition, DRGraph is equivalent to a force-directed based approach, if the gradient is split into one attractive force and M repulsive forces.
Therefore, it is feasible to employ conventional methods, such as simulated annealing~\cite{frick1994fast} to accelerate the convergence.
The simulated annealing technique randomly replaces system state (graph layout) into a new one with a high system energy to escape from local minima.
Meanwhile, constraints for dynamic graph layout (\eg, temporal coherence constraint \cite{brandes2011quantitative} $L_{temporal}=\sum_{t=1}^T\|Y^{t}-Y^{t-1}\|_2$) can be considered cost terms in the objective functions. 
Thus, the unified formulation provides the opportunity of applying varied functions or graph-related constraints into the formulation.

\textbf{Limitations.}
The sparse distance matrix and the negative sampling technique emphasize on preserving the local neighborhood structure while neglecting the global data structure.
Therefore, we employ a multi-level layout scheme to maintain the global structure.
There is no guarantee that all nodes are coarsened precisely because we generate coarse graphs randomly.
There are some edges with very large length (see the web-NotreDame dataset).
We aim to improve this in the future.

When the graph data size increases, it is not easy to achieve a good global layout.
Because moving nodes when there are full of nodes in the layout space is difficult.
%There are two solutions to tackle this issue: a good initialization and an early exaggeration.
tsNET* adopts the result of PMDS as the initial position of nodes.
DRGraph employs a multi-level scheme to find a good initialization with coarse graphs.
%We also employ the early exaggeration strategy used by $t$-SNE to find a good global structure in the early stages.
%the constant $\gamma$ is set to be $0.01$ for the coarsest graph and $\gamma = 0.1$ for the rest.

The data size influences the scalability of generating the node-link diagram.
When visualizing a graph data with millions of edges, the efficiency of visual exploration suffers from the limited screen space and possible visual clutter~\cite{7539318}.
We leverage the graph sampling technique~\cite{7864456} that randomly selects a subset of the edge set to capture the overall structure.
Other possible solutions are density-based visualization through splatting technique~\cite{van2003graphsplatting} and edge bundling~\cite{6634098,7539373}.

\section{Conclusions}\label{sec:conclusions}
In this paper, we present an efficient graph layout algorithm by enhancing the nonlinear dimensionality reduction method with several new techniques.
Our new method is feasible within $O(|V|+|E|+\text{TM}|V|)$ computational complexity and requires only $O(|V|+|E|)$ memory complexity.
Experimental results demonstrate that DRGraph achieves a significant acceleration and generates graph layouts of comparable quality to tsNET.
There are many future research directions. 
We plan to implement DRGraph in GPU by exploiting its parallelism and extend DRGraph for weighted graphs and dynamic graphs.
We expect to integrate DRGraph into a visual analysis system for large-scale graphs.

%The source code of DRGraph is available at https://github.com/DRGraphVIS/DRGraph.
%, and attribute networks. 

\acknowledgments{
This work is supported by National Natural Science Foundation of China (61772456, 61761136020).
}

\bibliographystyle{abbrv-doi}
%\bibliographystyle{abbrv-doi-narrow}
%\bibliographystyle{abbrv-doi-hyperref}
%\bibliographystyle{abbrv-doi-hyperref-narrow}

%\cleardoublepage
\bibliography{ms}
\end{document}

\maketitle

%\section*{Appendix: Parameter Sensitivity}
Tables \ref{table:parameter1}, \ref{table:parameter2}, \ref{table:parameter3}, \ref{table:parameter4}, and \ref{table:parameter5} show the parameter sensitivity.
Experiments on the com-Orkut and com-LiveJournal datasets are not feasible due to the enormous computational cost.
Some items are unfilled due to the memory limitation and the computational cost.

\textbf{The size of $k$-order nearest neighbors.}
Table \ref{table:parameter1} shows the neighborhood preservation accuracy (NP), the normalized stress, crosslessness and minimum
angle metrics with respect to the size of $k$-order nearest neighbors.
We find that NP drops when $k$ becomes large on almost all datasets.
The reason is that higher-order nearest neighbors contain dissimilar nodes, which are treated as positive nodes.
Thus, DRGraph places dissimilar nodes close to each other, resulting in a low layout quality.
When $k$ is small, DRGraph usually achieves higher results of crosslessness and minimum angle.
Besides, higher-order nearest neighbors cost large memory consumptions for keeping graph distances.
Thus, we choose the first-order nearest neighbors, which reduces the computational complexity to $O(|E|+|V|+\text{TM}|V|)$ and the memory complexity to $O(|E|+|V|)$.
$1$-order nearest neighbors is sufficient to provide locality properties, accelerate the computation of the node similarity, and meanwhile reduce the memory requirement.

\textbf{The number of negative samples $\text{M}$.}
Table \ref{table:parameter2} reports the layout quality with respect to the number of negative samples M.
We find that the performance slightly rises when the number of negative samples increases.
With a large subset of nodes, we can approximate the gradient accurately.
However, the computation complexity is linear with the number of negative samples.
To keep the balance between quality and efficiency, we choose the number of negative samples M=5.

\textbf{The weight of negative samples $\gamma$.}
Table \ref{table:parameter3} lists the layout quality with a varied $\gamma$.
$\gamma$ controls the value of repulsive forces of the gradient. 
A small value of $\gamma$ generates small repulsive forces, while nodes in the graph data are too close to each other and form local clusters.
We can see that the neighborhood preservation accuracy drops when $\gamma$ is very small.
A large value of $\gamma$ forces dissimilar nodes far away.
The edges are longer than the ideal length in the layout space, leading to bad stress quality.
The crosslessness and minimum angle metrics are stable for different choices of $\gamma$.
We choose a medium value of 0.1, which has a high neighborhood preservation accuracy and small stress.
DRGraph employs the early exaggeration technique to find a better initialization.
$\gamma$ is set to be 0.01 for the coarsest graph and $\gamma=0.1$ for the rest.
%A medium value of 0.01 has a good convergence and a good final quality, but the adaptive local speed achieves even better on convergence as well as on final quality. 

\textbf{The iteration number T.}
Table \ref{table:parameter4} presents the layout quality with respect to the iteration number T.
The accuracy of the neighborhood preservation becomes stable when the iteration number $\text{T}$ is large adequately.
Besides, the iteration number has little effect to the crosslessness and minimum angle metrics.
In practice, we choose the iteration number $\text{T}=400$ to accelerate the convergence and generate aesthetically pleasing results.

\begin{figure}[!ht]
  \centering
  \includegraphics[width=0.99\columnwidth]{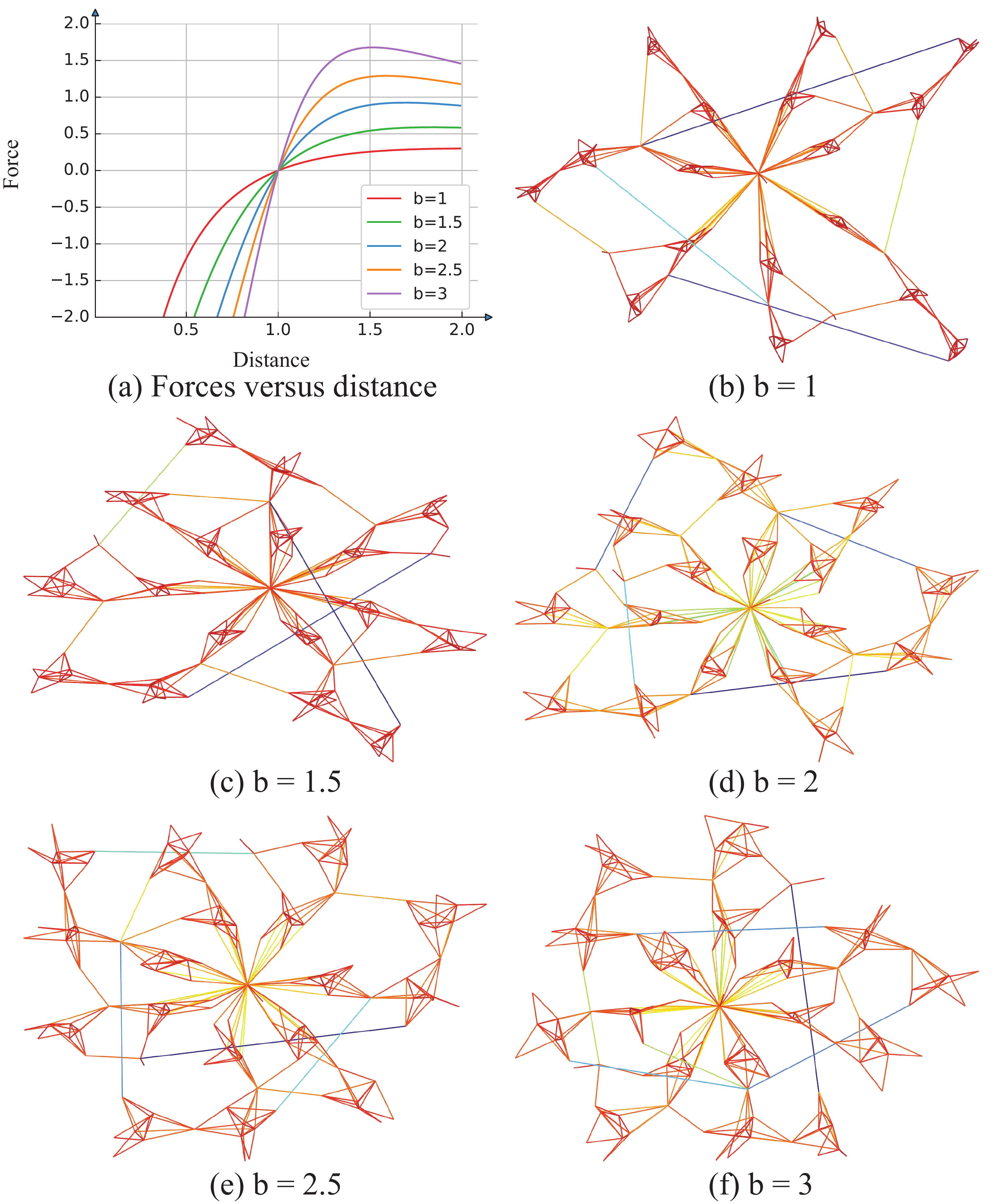}
  \caption{The effect of the $b$ parameter, which forces the edge to the ideal length.}
  \label{figure:b}
\end{figure}

\textbf{The effect of $\textbf{b}$.}
Figure \ref{figure:b} (a) illustrates the sum of the attractive force and the repulsive force (\ie, the gradient) with respect to $b$.
$b$ controls the value of the sum force without altering the ideal distance between nodes.
In general, a small $b$ value (\eg, $b = 1$) tends to put similar nodes close to others (see Figure \ref{figure:b} (b)).
As shown in Table \ref{table:parameter5}, a small value of $b$ would increase neighborhood preservation accuracy.
However, when $b$ is small, the layout distance varies greatly, resulting in a bad stress quality.
A very large $b$ value (\eg, $b = 3$) forces all edge lengths to be ideal but obfuscates the global structure (see Figure \ref{figure:b} (f)).
DRGraph achieves good stress and minimum angle quality but a low neighborhood preservation accuracy (see Table \ref{table:parameter5}) when $b$ is large.
In our implementation, we choose $b=3$ when the input graph is a grid graph (\eg, grid17), in which all edges have the same length.
For 3D meshes (\eg, G65) (\eg, G65) and large graphs (\eg, web-NotreDame), it is intractable to preserve all edge lengths of a manifold in the two-dimensional layout space. 
Therefore, we set $b=1$ to preserve the neighborhood identity.
For other graphs, we choose $b=2$, which works well in preserving both local and global structures.

\begin{table*}[]
\setlength{\tabcolsep}{3.2pt}
\centering
\caption{The layout quality with respect to the $k$-order nearest neighbors.}
\label{table:parameter1}
\scalebox{0.85}{
\begin{tabular}{@{}l|lllll|lllll|lllll|lllll@{}}
\toprule
$k$ & 1 & 2 & 3 & 4 & 5 & 1 & 2 & 3 & 4 & 5  & 1 & 2 & 3 & 4 & 5 & 1 & 2 & 3 & 4 & 5 \\ \midrule
 & \multicolumn{5}{c|}{NP} & \multicolumn{5}{c|}{stress} & \multicolumn{5}{c|}{crosslessness} & \multicolumn{5}{c}{min\_angle} \\\midrule
dwt\_72 & .8811 & .9151 & .8751 & .8663 & .8271 & .0428 & .0422 & .0480 & .0476 & .0470 & .9788 & 1.000 & .9878 & .9914 & .9878 & .9304 & .8888 & .8334 & .8076 & .7964 \\
lesmis & .6593 & .6852 & .5633 & .5111 & .5074 & .1255 & .1194 & .1526 & .1641 & .1737 & .8298 & .7963 & .7693 & .7564 & .7521 & .4201 & .3689 & .3601 & .3651 & .3580 \\
can\_96 & .6490 & .6197 & .5826 & .5607 & .5511 & .0881 & .0969 & .0985 & .1024 & .1053 & .9342 & .9319 & .9123 & .8917 & .8915 & .3286 & .2251 & .1775 & .1842 & .1748 \\
rajat11 & .7010 & .6538 & .5933 & .5665 & .5515 & .0989 & .0968 & .1012 & .0993 & .1023 & .9286 & .9052 & .8973 & .8918 & .8878 & .2069 & .1876 & .1872 & .1867 & .2057 \\
jazz & .7738 & .7538 & .6663 & .6394 & .6707 & .1368 & .1729 & .2194 & .2341 & .2164 & .7655 & .6837 & .6410 & .6174 & .6478 & .0809 & .0807 & .0748 & .0736 & .0745 \\
visbrazil & .4802 & .4561 & .3509 & .3157 & .2946 & .0833 & .0858 & .0993 & .0989 & .1008 & .9282 & .9027 & .8776 & .8710 & .8690 & .6600 & .6266 & .6206 & .6321 & .6136 \\
grid17 & .8505 & .7672 & .7331 & .7101 & .6985 & .0151 & .0171 & .0193 & .0216 & .0228 & 1.000 & .9980 & .9963 & .9904 & .9895 & .8547 & .8008 & .7480 & .7013 & .6763 \\
mesh3e1 & .9941 & .9035 & .8184 & .7839 & .7606 & .0042 & .0071 & .0116 & .0144 & .0159 & .9992 & .9964 & .9862 & .9815 & .9784 & .8138 & .6396 & .5026 & .4452 & .4087 \\
netscience & .6655 & .6453 & .5211 & .4423 & .4052 & .0965 & .0853 & .0819 & .0819 & .0823 & .9534 & .9397 & .9259 & .9183 & .9102 & .3772 & .3398 & .3255 & .3127 & .3065 \\
dwt\_419 & .7550 & .7256 & .6885 & .6665 & .6544 & .0189 & .0213 & .0234 & .0256 & .0268 & .9553 & .9512 & .9488 & .9463 & .9451 & .2078 & .1688 & .1701 & .1649 & .1609 \\
price\_1000 & .5567 & .4214 & .2796 & .1842 & .1389 & .1517 & .1311 & .1238 & .1224 & .1269 & .9888 & .9587 & .9343 & .9071 & .8834 & .8560 & .8090 & .8082 & .7966 & .7894 \\
dwt\_1005 & .5949 & .5629 & .5203 & .4919 & .4729 & .0399 & .0386 & .0300 & .0315 & .0373 & .9651 & .9637 & .9608 & .9575 & .9545 & .3325 & .2558 & .2065 & .1885 & .1899 \\
cage8 & .2963 & .2505 & .2106 & .1800 & .1629 & .1495 & .1453 & .1461 & .1517 & .1591 & .8962 & .8843 & .8685 & .8478 & .8375 & .1496 & .1345 & .1267 & .1142 & .1110 \\
bcsstk09 & .8865 & .8251 & .7580 & .7340 & .7171 & .0168 & .0202 & .0248 & .0268 & .0286 & .9566 & .9550 & .9513 & .9494 & .9478 & .0638 & .0846 & .0739 & .0702 & .0735 \\
block\_2000 & .3024 & .2755 & .2463 & .2000 & .1621 & .1805 & .1762 & .1607 & .1551 & .1605 & .8910 & .8706 & .8686 & .8630 & .8404 & .1172 & .1021 & .1021 & .0975 & .0936 \\
sierpinski3d & .5768 & .5596 & .5137 & .5098 & .4855 & .0754 & .0658 & .0854 & .0673 & .0738 & .9844 & .9828 & .9790 & .9781 & .9760 & .2394 & .2200 & .1906 & .1886 & .1864 \\
CA-GrQc & .1735 & .1543 & .1039 & .0712 & .0456 & .1894 & .1655 & .1555 & .1581 & .1742 & .9264 & .9160 & .8984 & .8740 & .8315 & .4139 & .3881 & .3691 & .3561 & .3413 \\
EVA & .7145 & .5866 & .4865 & .4216 & .3663 & .1529 & .2082 & .1802 & .1453 & .1390 & .9850 & .9589 & .9506 & .9381 & .9332 & .9319 & .9124 & .9142 & .9129 & .9125 \\
3elt & .6441 & .5872 & .5717 & .5596 & .5436 & .0612 & .0651 & .0722 & .0713 & .0744 & .9922 & .9930 & .9931 & .9920 & .9914 & .5037 & .4362 & .3813 & .3544 & .3283 \\
us\_powergrid & .4616 & .4245 & .3662 & .3248 & .3008 & .1176 & .1194 & .1104 & .1445 & .1228 & .9896 & .9872 & .9844 & .9820 & .9805 & .6898 & .6428 & .6147 & .5966 & .5873 \\
G65 & .2692 & .2518 & .2531 & .2556 & .2506 & .1208 & .1148 & .1173 & .1202 & .1184 & .9890 & .9886 & .9886 & .9886 & .9883 & .6388 & .5737 & .5521 & .5356 & .5212 \\
fe\_4elt2 & .5702 & .5350 & .5199 & .4764 & .4805 & .0722 & .0726 & .0748 & .0650 & .0632 & .9948 & .9945 & .9948 & .9936 & .9937 & .5024 & .4168 & .3461 & .3036 & .2871 \\
bcsstk31 & .3660 & .3953 & .3736 & .3564 & .3413 & .0624 & .0639 & .0671 & .0561 & .0589 & .9827 & .9822 & .9810 & .9800 & .9790 & .0238 & .0242 & .0245 & .0241 & .0235 \\
venkat50 & .6506 & .6069 & .5679 & .5484 & .5120 & .0593 & .0638 & .0635 & .0613 & .0653 & .9935 & .9933 & .9930 & .9929 & .9923 & .0209 & .0260 & .0267 & .0267 & .0252 \\
ship\_003 & .2002 & .2003 & .1894 & .1789 & .1638 & .0444 & .0450 & .0474 & .0484 & .0472 & .9843 & .9832 & .9820 & .9807 & .9789 & .0240 & .0278 & .0288 & .0289 & .0286 \\
troll & .2539 & .2579 & .2408 & .2242 & .2175 & .0960 & .0900 & .0933 & .0984 & .0917 & .9899 & .9891 & .9883 & .9875 & .9874 & .0091 & .0086 & .0086 & .0087 & .0090 \\
% torso3 & .1541 & .1488 & .1478 & .1483 & .1401 & .0725 & .0719 & .0819 & .1129 & .1233 & .9924 & .9921 & .9918 & .9917 & .9912 & .0562 & .0551 & .0519 & .0498 & .0473 \\
NotreDame & .4671 & .4192 & .2181 & (-) & (-) & .1331 & .1357 & .1380 & (-) & (-) & .9714 & .8099 & .8063 & (-) & (-) & .5603 & .5358 & .5424 & (-) & (-) \\
Flan\_1565 & .2067 & .2004 & .1898 & (-) & (-) & .0998 & .0867 & .0850 & (-) & (-) & (-) & (-) & (-) & (-) & (-) & .0081 & .0082 & .0073 & (-) & (-)
\\ \bottomrule
\end{tabular}
}
\end{table*}

\begin{table*}[]
\setlength{\tabcolsep}{3.2pt}
\centering
\caption{The layout quality with respect to the size of negative samples.}
\label{table:parameter2}
\scalebox{.85}{
\begin{tabular}{@{}l|lllll|lllll|lllll|lllll@{}}
\toprule
M & 1 & 3 & 5 & 7 & 10 & 1 & 3 & 5 & 7 & 10 & 1 & 3 & 5 & 7 & 10 & 1 & 3 & 5 & 7 & 10 \\\midrule
 & \multicolumn{5}{c|}{NP} & \multicolumn{5}{c|}{stress} & \multicolumn{5}{c|}{crosslessness} & \multicolumn{5}{c}{min\_angle} \\\midrule
dwt\_72 & .8889 & .8625 & .8473 & .8449 & .8357 & .0486 & .0475 & .0429 & .0493 & .0491 & 1.0000 & .9765 & .9804 & .9713 & .9654 & .9326 & .9145 & .9222 & .9163 & .9176 \\
lesmis & .6627 & .6481 & .6426 & .6349 & .6427 & .1225 & .1252 & .1273 & .1305 & .1319 & .8293 & .8303 & .8302 & .8317 & .8346 & .3999 & .4150 & .4206 & .4335 & .4277 \\
can\_96 & .5964 & .6321 & .6354 & .6280 & .6446 & .0800 & .0860 & .0872 & .0887 & .0916 & .9229 & .9335 & .9359 & .9237 & .9295 & .2229 & .2693 & .3167 & .3166 & .3521 \\
rajat11 & .7014 & .6945 & .6948 & .6945 & .6869 & .0944 & .0921 & .0964 & .1052 & .1009 & .9262 & .9296 & .9294 & .9281 & .9311 & .1862 & .1962 & .2102 & .2142 & .2335 \\
jazz & .7780 & .7759 & .7722 & .7718 & .7637 & .1439 & .1428 & .1360 & .1357 & .1373 & .7569 & .7621 & .7671 & .7664 & .7670 & .0774 & .0830 & .0809 & .0748 & .0790 \\
visbrazil & .4468 & .4710 & .4857 & .4910 & .4945 & .0814 & .0799 & .0835 & .0844 & .0881 & .9218 & .9268 & .9300 & .9285 & .9295 & .6353 & .6489 & .6531 & .6581 & .6628 \\
grid17 & .7408 & .8031 & .8424 & .8694 & .8979 & .0186 & .0158 & .0151 & .0147 & .0145 & .9948 & 1.000 & 1.000 & 1.000 & 1.000 & .7736 & .8342 & .8548 & .8669 & .8788 \\
mesh3e1 & .8991 & .9823 & .9947 & .9962 & .9966 & .0086 & .0049 & .0041 & .0036 & .0036 & .9937 & .9979 & 1.000 & 1.000 & 1.000 & 1.000 & .7891 & .8153 & .8204 & .8317 \\
netscience & .6277 & .6575 & .6647 & .6666 & .6603 & .0906 & .0931 & .0962 & .0943 & .1081 & .9497 & .9522 & .9535 & .9537 & .9533 & .3537 & .3760 & .3707 & .3840 & .3825 \\
dwt\_419 & .7464 & .7295 & .7543 & .7555 & .7500 & .0195 & .0271 & .0191 & .0192 & .0197 & .9548 & .9543 & .9552 & .9554 & .9558 & .1946 & .1818 & .2039 & .2108 & .2167 \\
price\_1000 & .4764 & .5343 & .5498 & .5595 & .5747 & .1394 & .1483 & .1488 & .1535 & .1549 & .9817 & .9860 & .9893 & .9905 & .9916 & .8461 & .8548 & .8563 & .8607 & .8639 \\
dwt\_1005 & .5425 & .5887 & .6094 & .6008 & .6149 & .0665 & .0575 & .0381 & .0371 & .0378 & .9625 & .9653 & .9670 & .9656 & .9670 & .2834 & .3248 & .3411 & .3498 & .3492 \\
cage8 & .2303 & .2756 & .2916 & .3174 & .3343 & .1384 & .1453 & .1467 & .1497 & .1541 & .8853 & .8920 & .8947 & .8976 & .8988 & .1500 & .1500 & .1475 & .1505 & .1486 \\
bcsstk09 & .8361 & .8752 & .8866 & .8922 & .8909 & .0193 & .0172 & .0168 & .0166 & .0167 & .9558 & .9564 & .9566 & .9566 & .9567 & .0770 & .0660 & .0627 & .0633 & .0650 \\
block\_2000 & .2767 & .2931 & .3028 & .3075 & .3151 & .1724 & .1738 & .1790 & .1788 & .1802 & .8812 & .8872 & .8911 & .8927 & .8952 & .1083 & .1139 & .1177 & .1195 & .1210 \\
sierpinski3d & .5085 & .5637 & .5645 & .5745 & .5903 & .0933 & .0768 & .0761 & .0816 & .0853 & .9807 & .9837 & .9838 & .9845 & .9849 & .2097 & .2234 & .2346 & .2421 & .2493 \\
CA-GrQc & .1306 & .1591 & .1730 & .1854 & .2003 & .1903 & .1873 & .1867 & .1890 & .1875 & .9202 & .9253 & .9260 & .9270 & .9285 & .4109 & .4111 & .4162 & .4172 & .4170 \\
EVA & .6569 & .7010 & .7188 & .7265 & .7340 & .1477 & .1541 & .1492 & .1513 & .1563 & .9807 & .9840 & .9854 & .9856 & .9866 & .9309 & .9306 & .9308 & .9319 & .9324 \\
3elt & .5776 & .5924 & .6438 & .6450 & .6596 & .0683 & .0646 & .0670 & .0637 & .0641 & .9926 & .9927 & .9927 & .9931 & .9921 & .4582 & .4816 & .5048 & .5001 & .5124 \\
us\_powergrid & .4152 & .4488 & .4642 & .4732 & .4833 & .1387 & .1284 & .1175 & .1124 & .1178 & .9879 & .9893 & .9895 & .9899 & .9899 & .6794 & .6875 & .6939 & .6994 & .6988 \\
G65 & .2422 & .2587 & .2615 & .2625 & .2639 & .1189 & .1165 & .1158 & .1140 & .1157 & .9877 & .9887 & .9887 & .9886 & .9884 & .5927 & .6022 & .6075 & .6077 & .6152 \\
fe\_4elt2 & .4739 & .5771 & .5867 & .5969 & .5952 & .0828 & .0751 & .0743 & .0712 & .0718 & .9936 & .9951 & .9945 & .9952 & .9949 & .4364 & .4973 & .5057 & .5229 & .5222 \\
bcsstk31 & .3580 & .3780 & .3920 & .3866 & .4014 & .0603 & .0657 & .0640 & .0787 & .0773 & .9827 & .9832 & .9836 & .9832 & .9837 & .0239 & .0248 & .0250 & .0244 & .0251 \\
venkat50 & .5364 & .6040 & .6382 & .6473 & .6408 & .1188 & .0759 & .0572 & .0721 & .0816 & .9926 & .9932 & .9934 & .9934 & .9931 & .0197 & .0205 & .0209 & .0210 & .0205 \\
ship\_003 & .1818 & .1944 & .2000 & .2031 & .2084 & .0415 & .0472 & .0437 & .0407 & .0463 & .9836 & .9841 & .9843 & .9843 & .9844 & .0247 & .0244 & .0241 & .0240 & .0236 \\
troll & .2419 & .2488 & .2523 & .2585 & .2624 & .0997 & .1061 & .1007 & .1019 & .1113 & .9897 & .9899 & .9899 & .9900 & .9901 & .0092 & .0092 & .0092 & .0093 & .0093 \\
torso3 & .1335 & .1416 & .1481 & .1413 & .1477 & .1275 & .0899 & .0734 & .0979 & .1333 & .9916 & .9918 & .9922 & .9921 & .9920 & .0540 & .0552 & .0560 & .0582 & .0576 \\
NotreDame & .4365 & .4569 & .4650 & .4779 & .4786 & .1318 & .1270 & .1286 & .1327 & .1305 & .9713 & .9714 & .9714 & .9714 & .9713 & .5571 & .5605 & .5606 & .5617 & .5612 \\
Flan\_1565 & .1812 & .1999 & .2040 & .2102 & .2107 & .0907 & .0924 & .0922 & .0840 & .0866 & (-) & (-) & (-) & (-) & (-) & .0074 & .0077 & .0081 & .0084 & .0084 \\\bottomrule
\end{tabular}
}
\end{table*}

\begin{table*}[]
\setlength{\tabcolsep}{3.2pt}
\centering
\caption{The layout quality with respect to $\gamma$.}
\label{table:parameter3}
\scalebox{.85}{
\begin{tabular}{@{}l|lllll|lllll|lllll|lllll@{}}
\toprule
$\gamma$ & 0.01 & 0.05 & 0.1 & 0.2 & 1& 0.01 & 0.05 & 0.1 & 0.2 & 1& 0.01 & 0.05 & 0.1 & 0.2 & 1& 0.01 & 0.05 & 0.1 & 0.2 & 1 \\\midrule
 & \multicolumn{5}{c|}{NP} & \multicolumn{5}{c|}{stress} & \multicolumn{5}{c|}{crosslessness} & \multicolumn{5}{c}{min\_angle} \\\midrule
dwt\_72 & .7951 & .8217 & .8834 & .8981 & .9159 & .0568 & .0551 & .0488 & .0432 & .0514 & .9680 & .9807 & .9852 & .9880 & .9814 & .9360 & .9296 & .9213 & .9158 & .9169 \\
lesmis & .6689 & .6534 & .6552 & .6411 & .6361 & .1120 & .1272 & .1234 & .1365 & .1451 & .8353 & .8354 & .8358 & .8352 & .8358 & .4006 & .4090 & .4183 & .4281 & .4261 \\
can\_96 & .6015 & .6172 & .6348 & .6371 & .6515 & .0775 & .0839 & .0888 & .0915 & .0983 & .9358 & .9255 & .9295 & .9290 & .9398 & .3220 & .3231 & .3215 & .3148 & .2981 \\
rajat11 & .6704 & .7004 & .7031 & .7008 & .6958 & .0767 & .1138 & .0996 & .1195 & .1413 & .9310 & .9278 & .9300 & .9279 & .9289 & .2166 & .2121 & .2082 & .2009 & .2110 \\
jazz & .7923 & .7788 & .7700 & .7708 & .7557 & .1296 & .1374 & .1365 & .1255 & .1488 & .7614 & .7641 & .7643 & .7639 & .7661 & .0788 & .0777 & .0816 & .0787 & .0805 \\
visbrazil & .4623 & .4724 & .4841 & .4979 & .5100 & .0777 & .0808 & .0840 & .0891 & .1039 & .9271 & .9292 & .9276 & .9298 & .9320 & .6464 & .6540 & .6513 & .6519 & .6700 \\
grid17 & .8386 & .8495 & .8579 & .8564 & .8636 & .0155 & .0150 & .0149 & .0149 & .0147 & 1.000 & 1.000 & 1.000 & 1.000 & 1.000 & .8681 & .8641 & .8647 & .8607 & .8736 \\
mesh3e1 & .9757 & .9929 & .9956 & .9971 & .9913 & .0076 & .0044 & .0040 & .0037 & .0040 & .9979 & .9985 & 1.000 & 1.000 & 1.000 & .8224 & .8185 & .8122 & .8118 & .8156 \\
netscience & .6293 & .6585 & .6684 & .6725 & .6888 & .0959 & .0969 & .0940 & .0963 & .1264 & .9525 & .9526 & .9519 & .9540 & .9542 & .3864 & .3744 & .3732 & .3605 & .3731 \\
dwt\_419 & .7555 & .7570 & .7585 & .7541 & .7261 & .0179 & .0182 & .0188 & .0200 & .0311 & .9556 & .9556 & .9554 & .9553 & .9545 & .2162 & .2025 & .2071 & .1952 & .1661 \\
price\_1000 & .5144 & .5453 & .5493 & .5651 & .5855 & .1430 & .1409 & .1477 & .1612 & .1660 & .9867 & .9885 & .9891 & .9899 & .9891 & .8565 & .8590 & .8562 & .8523 & .8518 \\
dwt\_1005 & .5867 & .6080 & .5900 & .5950 & .5965 & .0487 & .0382 & .0413 & .0516 & .0566 & .9662 & .9671 & .9659 & .9657 & .9662 & .3550 & .3583 & .3412 & .3361 & .3194 \\
cage8 & .2224 & .2670 & .2955 & .3218 & .3560 & .1332 & .1423 & .1481 & .1520 & .1605 & .8863 & .8931 & .8956 & .8980 & .9003 & .1564 & .1507 & .1483 & .1474 & .1440 \\
bcsstk09 & .8898 & .8878 & .8896 & .8915 & .8826 & .0167 & .0167 & .0167 & .0168 & .0174 & .9566 & .9566 & .9566 & .9566 & .9568 & .0574 & .0619 & .0623 & .0643 & .0643 \\
block\_2000 & .2735 & .2934 & .3020 & .3110 & .3237 & .1678 & .1729 & .1757 & .1820 & .1865 & .8849 & .8882 & .8911 & .8931 & .8966 & .1126 & .1158 & .1158 & .1185 & .1237 \\
sierpinski3d & .5561 & .5730 & .5717 & .5742 & .5676 & .0850 & .0791 & .0773 & .0827 & .0530 & .9843 & .9845 & .9841 & .9844 & .9840 & .2434 & .2409 & .2378 & .2354 & .2329 \\
CA-GrQc & .1349 & .1570 & .1754 & .1936 & .2398 & .1801 & .1863 & .1862 & .1900 & .1938 & .9213 & .9244 & .9267 & .9283 & .9301 & .4210 & .4130 & .4168 & .4155 & .4181 \\
EVA & .6865 & .6963 & .7145 & .7348 & .7687 & .1451 & .1464 & .1506 & .1545 & .1545 & .9823 & .9844 & .9848 & .9856 & .9866 & .9320 & .9316 & .9310 & .9312 & .9319 \\
3elt & .5886 & .5867 & .6455 & .6353 & .6305 & .0683 & .0646 & .0631 & .0765 & .0733 & .9933 & .9924 & .9919 & .9941 & .9929 & .4943 & .5035 & .5061 & .5179 & .4867 \\
us\_powergrid & .4405 & .4585 & .4673 & .4669 & .4737 & .1150 & .1107 & .1145 & .1204 & .1222 & .9894 & .9895 & .9896 & .9896 & .9894 & .6908 & .6948 & .6937 & .6957 & .6942 \\
G65 & .2427 & .2621 & .2555 & .2702 & .2771 & .1160 & .1195 & .1129 & .1186 & .1157 & .9889 & .9891 & .9886 & .9888 & .9886 & .5931 & .6245 & .6098 & .6218 & .6215 \\
fe\_4elt2 & .5017 & .5763 & .5831 & .5658 & .5796 & .0609 & .0710 & .0712 & .0787 & .0879 & .9945 & .9951 & .9947 & .9945 & .9949 & .4467 & .5087 & .5127 & .5115 & .5136 \\
bcsstk31 & .3741 & .3821 & .3847 & .3976 & .3918 & .0567 & .0611 & .0652 & .0660 & .0688 & .9838 & .9836 & .9835 & .9835 & .9830 & .0226 & .0245 & .0249 & .0247 & .0232 \\
venkat50 & .5822 & .6402 & .6398 & .6478 & .6224 & .0842 & .0718 & .0602 & .0761 & .0935 & .9933 & .9934 & .9933 & .9934 & .9932 & .0178 & .0201 & .0203 & .0206 & .0198 \\
ship\_003 & .1795 & .1946 & .1997 & .2017 & .2126 & .0463 & .0896 & .0443 & .0455 & .0515 & .9840 & .9842 & .9843 & .9842 & .9843 & .0241 & .0240 & .0240 & .0238 & .0225 \\
troll & .2380 & .2497 & .2548 & .2576 & .2669 & .0944 & .0845 & .0969 & .0962 & .0975 & .9900 & .9900 & .9900 & .9900 & .9901 & .0081 & .0089 & .0091 & .0092 & .0093 \\
torso3 & .1392 & .1441 & .1486 & .1508 & .1465 & .0708 & .0773 & .0716 & .0846 & .0850 & .9921 & .9921 & .9921 & .9921 & .9920 & .0440 & .0540 & .0558 & .0581 & .0611 \\
web-NotreDame & .5051 & .4777 & .4704 & .4532 & .4237 & .1264 & .1250 & .1288 & .1282 & .1316 & .9715 & .9713 & .9715 & .9717 & .9716 & .5622 & .5621 & .5608 & .5589 & .5531 \\
Flan\_1565 & .1875 & .2032 & .2043 & .2043 & .2056 & .0900 & .0896 & .0924 & .9599 & .0952 & (-) & (-) & (-) & (-) & (-) & .0064 & .0079 & .0081 & .0082 & .0084
\\\bottomrule
\end{tabular}
}
\end{table*}

\begin{table*}[]
\setlength{\tabcolsep}{3.2pt}
\centering
\caption{The layout quality with respect to the iteration number T.}
\label{table:parameter4}
\scalebox{.85}{
\begin{tabular}{@{}l|lllll|lllll|lllll|lllll@{}}
\toprule
T & 100 & 200 & 300 & 400 & 500 & 100 & 200 & 300 & 400 & 500 & 100 & 200 & 300 & 400 & 500 & 100 & 200 & 300 & 400 & 500 \\\midrule
 & \multicolumn{5}{c|}{NP} & \multicolumn{5}{c|}{stress} & \multicolumn{5}{c|}{crosslessness} & \multicolumn{5}{c}{min\_angle} \\\midrule
dwt\_72 & .6705 & .7446 & .8388 & .8808 & .9098 & .1104 & .1021 & .0740 & .0476 & .0370 & .9577 & .9644 & .9710 & .9795 & .9827 & .8658 & .8949 & .9191 & .9229 & .9337 \\
lesmis & .6156 & .6355 & .6513 & .6538 & .6468 & .1466 & .1342 & .1276 & .1284 & .1290 & .8285 & .8303 & .8298 & .8319 & .8323 & .4058 & .4183 & .4266 & .4199 & .4274 \\
can\_96 & .5988 & .6479 & .6577 & .6551 & .6509 & .1061 & .0930 & .0911 & .0865 & .0845 & .9073 & .9315 & .9344 & .9342 & .9346 & .1843 & .2677 & .2995 & .3197 & .3234 \\
rajat11 & .6555 & .6837 & .6827 & .6887 & .7006 & .1194 & .1100 & .0998 & .0991 & .0982 & .9230 & .9242 & .9283 & .9271 & .9295 & .2019 & .2079 & .2038 & .2104 & .2133 \\
jazz & .7476 & .7606 & .7683 & .7762 & .7713 & .1593 & .1578 & .1398 & .1366 & .1332 & .7485 & .7568 & .7627 & .7643 & .7675 & .0797 & .0796 & .0769 & .0814 & .0817 \\
visbrazil & .4370 & .4675 & .4809 & .4829 & .4930 & .0942 & .0865 & .0854 & .0823 & .0826 & .9218 & .9247 & .9275 & .9301 & .9266 & .6510 & .6635 & .6587 & .6548 & .6557 \\
grid17 & .7511 & .8022 & .8321 & .8473 & .8527 & .0183 & .0160 & .0154 & .0150 & .0149 & .9967 & 1.000 & 1.000 & 1.000 & 1.000 & .7885 & .8339 & .8472 & .8544 & .8617 \\
mesh3e1 & .8943 & .9705 & .9878 & .9934 & .9960 & .0108 & .0057 & .0044 & .0040 & .0039 & .9945 & .9976 & .9987 & 1.000 & 1.000 & .6969 & .7646 & .7992 & .8128 & .8229 \\
netscience & .6113 & .6527 & .6606 & .6642 & .6654 & .1015 & .0985 & .0942 & .0957 & .1032 & .9484 & .9521 & .9527 & .9531 & .9536 & .3537 & .3629 & .3713 & .3747 & .3798 \\
dwt\_419 & .7129 & .7077 & .7228 & .7563 & .7583 & .0234 & .0346 & .0272 & .0190 & .0189 & .9534 & .9527 & .9537 & .9543 & .9549 & .1759 & .1894 & .1874 & .1993 & .2000 \\
price\_1000 & .4723 & .5271 & .5505 & .5526 & .5560 & .1566 & .1535 & .1502 & .1486 & .1484 & .9822 & .9867 & .9889 & .9888 & .9890 & .8446 & .8501 & .8567 & .8564 & .8562 \\
dwt\_1005 & .5462 & .5744 & .5928 & .5931 & .5917 & .0607 & .0657 & .0382 & .0417 & .0403 & .9631 & .9640 & .9666 & .9658 & .9656 & .2526 & .3025 & .3340 & .3481 & .3456 \\
cage8 & .2738 & .2848 & .2968 & .2969 & .2958 & .1502 & .1485 & .1471 & .1481 & .1470 & .8878 & .8927 & .8953 & .8964 & .8959 & .1410 & .1418 & .1455 & .1513 & .1508 \\
bcsstk09 & .8253 & .8643 & .8789 & .8879 & .8936 & .0202 & .0178 & .0171 & .0167 & .0165 & .9553 & .9564 & .9565 & .9566 & .9566 & .0798 & .0710 & .0686 & .0648 & .0603 \\
block\_2000 & .2887 & .2965 & .3005 & .3025 & .3032 & .1832 & .1775 & .1778 & .1781 & .1800 & .8829 & .8876 & .8894 & .8908 & .8917 & .1105 & .1180 & .1186 & .1171 & .1174 \\
sierpinski3d & .4956 & .5503 & .5737 & .5743 & .5890 & .0887 & .0844 & .0771 & .0752 & .0680 & .9783 & .9823 & .9840 & .9843 & .9850 & .1934 & .2145 & .2311 & .2335 & .2436 \\
CA-GrQc & .1566 & .1697 & .1730 & .1746 & .1761 & .1971 & .1910 & .1880 & .1874 & .1847 & .9229 & .9253 & .9260 & .9261 & .9268 & .4038 & .4106 & .4109 & .4153 & .4140 \\
EVA & .6294 & .6840 & .7022 & .7165 & .7187 & .1474 & .1462 & .1550 & .1527 & .1515 & .9812 & .9833 & .9844 & .9858 & .9851 & .9280 & .9298 & .9306 & .9318 & .9318 \\
3elt & .5409 & .5941 & .6024 & .6467 & .6423 & .0782 & .1000 & .0744 & .0627 & .0685 & .9912 & .9917 & .9918 & .9939 & .9939 & .3856 & .4557 & .4821 & .5114 & .5086 \\
us\_powergrid & .3990 & .4419 & .4584 & .4610 & .4642 & .1275 & .1094 & .1132 & .1127 & .1104 & .9867 & .9886 & .9894 & .9895 & .9895 & .6690 & .6855 & .6909 & .6920 & .6954 \\
G65 & .2620 & .2642 & .2520 & .2577 & .2562 & .1163 & .1158 & .1140 & .1161 & .1115 & .9882 & .9885 & .9882 & .9883 & .9885 & .5726 & .6083 & .6005 & .6124 & .5874 \\
fe\_4elt2 & .5015 & .5633 & .5765 & .5829 & .5980 & .0773 & .0785 & .0821 & .0787 & .0704 & .9935 & .9947 & .9946 & .9946 & .9953 & .3954 & .4917 & .4885 & .5083 & .5198 \\
bcsstk31 & .3499 & .3754 & .3718 & .3808 & .3897 & .0903 & .0816 & .0657 & .0646 & .0601 & .9810 & .9824 & .9828 & .9833 & .9839 & .0249 & .0251 & .0247 & .0243 & .0243 \\
venkat50 & .5171 & .6066 & .6327 & .6408 & .6418 & .0750 & .0725 & .0639 & .0590 & .0661 & .9919 & .9931 & .9933 & .9934 & .9933 & .0259 & .0243 & .0228 & .0209 & .0198 \\
ship\_003 & .1916 & .1973 & .1989 & .2000 & .2008 & .0845 & .0548 & .0458 & .0432 & .0425 & .9830 & .9838 & .9841 & .9843 & .9844 & .0253 & .0247 & .0244 & .0243 & .0240 \\
troll & .2413 & .2527 & .2570 & .2527 & .2535 & .0956 & .0901 & .0912 & .0948 & .0957 & .9891 & .9898 & .9899 & .9899 & .9900 & .0106 & .0102 & .0096 & .0092 & .0088 \\
torso3 & .1359 & .1395 & .1489 & .1489 & .1492 & .1159 & .0850 & .0731 & .0722 & .0729 & .9912 & .9917 & .9923 & .9922 & .9923 & .0635 & .0626 & .0578 & .0561 & .0546 \\
web-NotreDame & .3581 & .4160 & .4487 & .4680 & .4815 & .1348 & .1269 & .1230 & .1283 & .1272 & .9701 & .9709 & .9712 & .9715 & .9716 & .5552 & .5573 & .5597 & .5608 & .5609 \\
Flan\_1565 & .1852 & .2041 & .2046 & .2045 & .2026 & .1050 & .0817 & .0940 & .0936 & .0889 & (-) & (-) & (-) & (-) & (-) & .0085 & .0089 & .0082 & .0079 & .0079
\\\bottomrule
\end{tabular}
}
\end{table*}

\begin{table*}[!t]
\setlength{\tabcolsep}{3.2pt}
\centering
\caption{The layout quality with respect to the parameter $b$.}
\label{table:parameter5}
\scalebox{0.85}{
\begin{tabular}{@{}l|lllll|lllll|lllll|lllll@{}}
\toprule
$b$ & 1 & 1.5 & 2 & 2.5 & 3 & 1 & 1.5 & 2 & 2.5 & 3 & 1 & 1.5 & 2 & 2.5 & 3 & 1 & 1.5 & 2 & 2.5 & 3 \\ \midrule
 & \multicolumn{5}{c|}{NP} & \multicolumn{5}{c|}{stress} & \multicolumn{5}{c|}{crosslessness} & \multicolumn{5}{c}{min\_angle} \\\midrule
dwt\_72 & .7668 & .8259 & .8773 & .8385 & .7613 & .0836 & .0720 & .0449 & .0512 & .0670 & .9577 & .9771 & .9808 & .9827 & .9727 & .9292 & .9200 & .9281 & .9095 & .8991 \\
lesmis & .6512 & .6626 & .6564 & .6482 & .6245 & .1657 & .1319 & .1227 & .1276 & .1306 & .8391 & .8365 & .8301 & .8296 & .8279 & .4272 & .4260 & .4251 & .4184 & .4191 \\
can\_96 & .6055 & .6312 & .6496 & .6421 & .6289 & .0894 & .0889 & .0866 & .0905 & .0894 & .9399 & .9350 & .9333 & .9348 & .9300 & .2157 & .3061 & .3129 & .3228 & .3222 \\
rajat11 & .6870 & .7012 & .7026 & .6735 & .6633 & .1114 & .1025 & .0970 & .0999 & .1069 & .9292 & .9308 & .9281 & .9272 & .9264 & .2318 & .2093 & .2051 & .1860 & .2006 \\
jazz & .7843 & .7755 & .7744 & .7658 & .7564 & .1626 & .1406 & .1353 & .1408 & .1457 & .7704 & .7698 & .7672 & .7639 & .7609 & .0782 & .0781 & .0806 & .0836 & .0775 \\
visbrazil & .5180 & .4935 & .4806 & .4711 & .4599 & .1107 & .0878 & .0849 & .0865 & .0856 & .9344 & .9309 & .9269 & .9266 & .9243 & .6903 & .6640 & .6529 & .6466 & .6457 \\
grid17 & .7049 & .7947 & .8472 & .8544 & .8540 & .0262 & .0164 & .0150 & .0148 & .0149 & 1.000 & 1.000 & 1.000 & 1.000 & 1.000 & .8102 & .8646 & .8724 & .8675 & .8613 \\
mesh3e1 & .8012 & .9387 & .9905 & .9947 & .9966 & .0175 & .0063 & .0040 & .0039 & .0040 & .9989 & 1.000 & 1.000 & .9987 & 1.000 & .5796 & .7499 & .8112 & .8180 & .8181 \\
netscience & .6892 & .6859 & .6603 & .6345 & .6161 & .1247 & .0990 & .0940 & .0938 & .0891 & .9560 & .9551 & .9528 & .9518 & .9508 & .3564 & .3798 & .3770 & .3682 & .3573 \\
dwt\_419 & .7033 & .7263 & .7501 & .7523 & .7195 & .0266 & .0189 & .0181 & .0187 & .0176 & .9544 & .9542 & .9545 & .9557 & .9537 & .1147 & .1641 & .2088 & .2093 & .2055 \\
price\_1000 & .6036 & .5893 & .5484 & .5080 & .4674 & .1782 & .1765 & .1507 & .1515 & .1455 & .9923 & .9908 & .9883 & .9862 & .9843 & .8589 & .8617 & .8553 & .8557 & .8505 \\
dwt\_1005 & .5080 & .5750 & .5823 & .5809 & .5939 & .0454 & .0462 & .0403 & .0427 & .0425 & .9649 & .9660 & .9653 & .9650 & .9658 & .1971 & .2972 & .3323 & .3384 & .3417 \\
cage8 & .3564 & .3252 & .2967 & .2824 & .2686 & .1651 & .1524 & .1469 & .1444 & .1432 & .9001 & .8986 & .8962 & .8936 & .8925 & .1448 & .1475 & .1494 & .1462 & .1511 \\
bcsstk09 & .7833 & .8567 & .8833 & .8880 & .8857 & .0272 & .0194 & .0174 & .0169 & .0168 & .9566 & .9569 & .9567 & .9567 & .9566 & .0727 & .0694 & .0640 & .0626 & .0631 \\
block\_2000 & .3207 & .3098 & .3020 & .2975 & .2926 & .1999 & .1863 & .1795 & .1732 & .1719 & .8980 & .8942 & .8904 & .8892 & .8876 & .1234 & .1185 & .1156 & .1137 & .1142 \\
sierpinski3d & .5257 & .5684 & .5746 & .5438 & .5354 & .0918 & .0761 & .0767 & .0874 & .0876 & .9844 & .9847 & .9842 & .9826 & .9822 & .1994 & .2259 & .2336 & .2302 & .2227 \\
CA-GrQc & .3081 & .2174 & .1736 & .1537 & .1407 & .2594 & .1997 & .1872 & .1839 & .1819 & .9324 & .9292 & .9262 & .9239 & .9224 & .4166 & .4202 & .4176 & .4101 & .4079 \\
EVA & .7957 & .7646 & .7118 & .6648 & .6231 & .1705 & .1536 & .1529 & .1533 & .1468 & .9899 & .9877 & .9845 & .9829 & .9815 & .9338 & .9334 & .9312 & .9294 & .9284 \\
3elt & .6432 & .6530 & .5504 & .5116 & .4885 & .0639 & .0689 & .0793 & .0839 & .0866 & .9927 & .9930 & .9902 & .9886 & .9877 & .5093 & .5549 & .5021 & .4164 & .3550 \\
us\_powergrid & .4629 & .4227 & .3502 & .2827 & .2491 & .1166 & .1171 & .1210 & .1239 & .1222 & .9896 & .9882 & .9864 & .9840 & .9829 & .6931 & .6892 & .6878 & .6424 & .6279 \\
G65 & .2511 & .2694 & .2615 & .2451 & .2188 & .1145 & .1138 & .1146 & .1173 & .1192 & .9882 & .9883 & .9876 & .9866 & .9849 & .6132 & .6352 & .6169 & .5470 & .5098 \\
fe\_4elt2 & .5836 & .5784 & .4979 & .4260 & .3756 & .0750 & .0879 & .0951 & .0940 & .1002 & .9949 & .9942 & .9925 & .9904 & .9885 & .5086 & .5576 & .4177 & .3256 & .2786 \\
bcsstk31 & .3893 & .3944 & .3788 & .3418 & .3416 & .0626 & .0617 & .0654 & .0785 & .0704 & .9836 & .9830 & .9826 & .9812 & .9809 & .0246 & .0286 & .0313 & .0321 & .0326 \\
venkat50 & .6431 & .6084 & .5222 & .4309 & .3802 & .0576 & .0668 & .0765 & .0862 & .1054 & .9932 & .9928 & .9916 & .9898 & .9881 & .0205 & .0251 & .0300 & .0294 & .0284 \\
ship\_003 & .1976 & .1979 & .1947 & .1877 & .1743 & .0474 & .0463 & .0426 & .0476 & .0509 & .9840 & .9838 & .9835 & .9827 & .9815 & .0243 & .0269 & .0291 & .0312 & .0321 \\
troll & .2548 & .2505 & .2365 & .2181 & .2022 & .0976 & .0914 & .1029 & .1143 & .0978 & .9900 & .9897 & .9890 & .9877 & .9866 & .0091 & .0109 & .0128 & .0133 & .0133 \\
torso3 & .1450 & .1376 & .1207 & .1074 & .0949 & .0749 & .0850 & .1053 & .1099 & .1438 & .9921 & .9919 & .9910 & .9895 & .9880 & .0556 & .0718 & .0724 & .0672 & .0643 \\
web-NotreDame & .4752 & .3458 & .2855 & .2533 & .2358 & .1411 & .1398 & .1355 & .1433 & .1384 & .9714 & .9691 & .9670 & .9654 & .9641 & .5604 & .5546 & .5499 & .5463 & .5455 \\
Flan\_1565 & .2077 & .1968 & .1575 & .1258 & .0976 & .0927 & .0949 & .1079 & .1019 & .1392 & (-) & (-) & (-) & (-) & (-) & .0082 & .0105 & .0101 & .0096 & .0093
\\ \bottomrule
\end{tabular}
}
\end{table*}

\begin{figure*}[!ht]
  \centering
  \includegraphics[width=1.99\columnwidth]{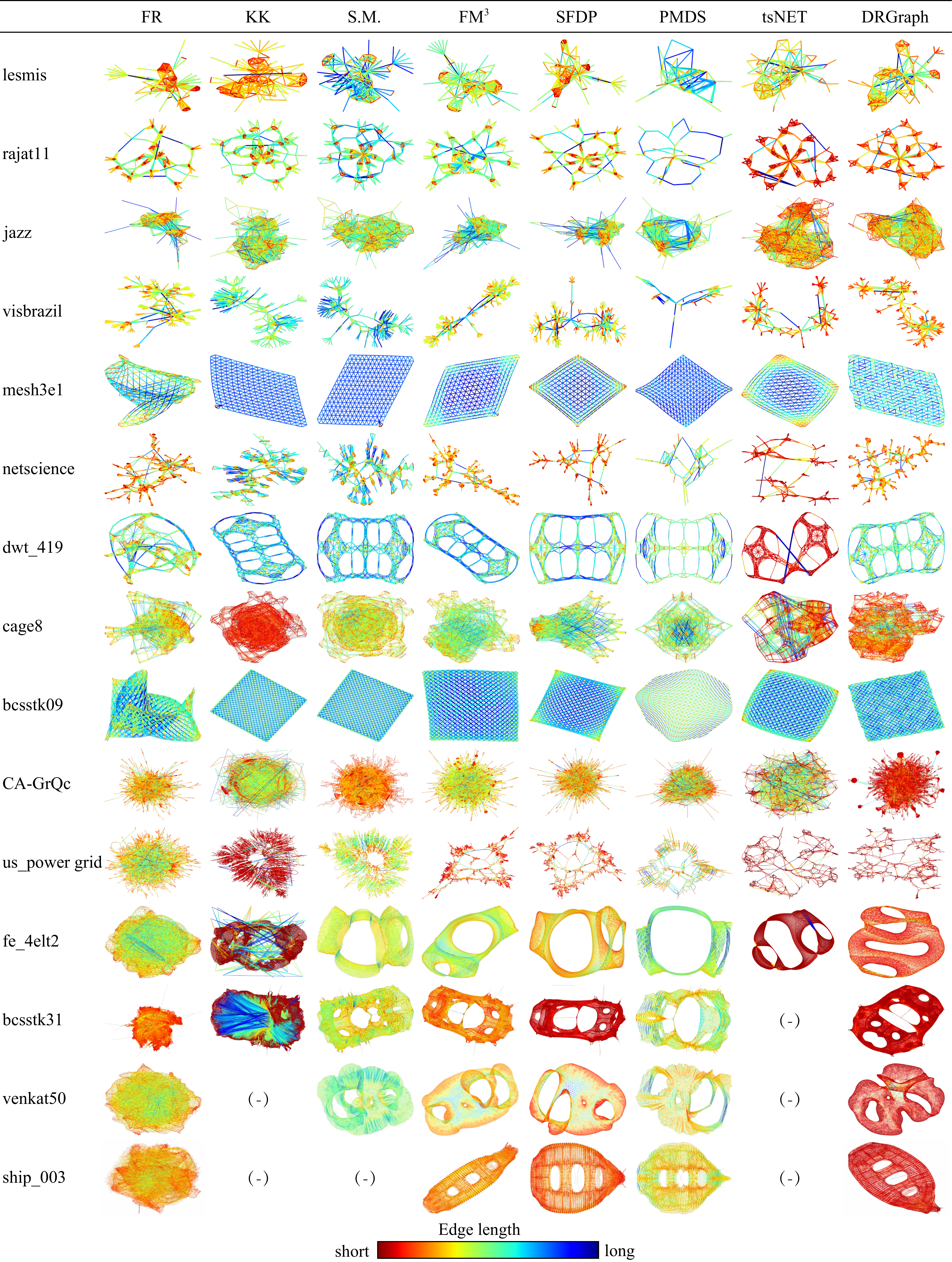}
  \caption{Visualizations of selected graph datasets using FR, KK, Stress Majorization (S.M.), $\text{FM}^3$, SFDP, PMDS, tsNET and DRGraph.
%   The edge length is encoded with the color map used in \cite{gansner2013maxent}. 
  }
  \label{figure:visualization2}
\end{figure*}

\bibliographystyle{abbrv-doi}
%\bibliographystyle{abbrv-doi-narrow}
%\bibliographystyle{abbrv-doi-hyperref}
%\bibliographystyle{abbrv-doi-hyperref-narrow}

%\cleardoublepage
%\bibliography{template}